\documentclass[a4paper,11pt]{article}
\pdfoutput=1 

\usepackage{jheppub} 

\usepackage[T1]{fontenc} 

\usepackage{soul}
\usepackage{mathrsfs}
\usepackage{amsmath}
\usepackage{subcaption}
\usepackage{amsfonts}
\usepackage{booktabs}
\usepackage{siunitx}
\usepackage{array}
\newcolumntype{M}[1]{>{\centering\arraybackslash}m{#1}}
\usepackage{verbatim}
\usepackage{float}
\RequirePackage{xcolor}
\usepackage{layouts}

\def\bea{\begin{eqnarray}}
\def\eea{\end{eqnarray}}
\def\be{\begin{equation}}
\def\ee{\end{equation}}

\newcommand{\fig}{Fig.~}
\newcommand{\figs}{Figs.~}
\newcommand{\eq}{Eq.~}
\newcommand{\eqs}{Eqs.~}
\newcommand{\se}{Sec.~}

\newcommand{\re}{Ref.~}

\newcommand{\app}{App.~}

\newcommand{\tab}{Table~}

\renewcommand{\Re}{\rm Re}

\newcommand{\Tr}{\mathrm{Tr}}

\newcommand{\mx}{\mathbf x}

\newcommand{\p}{\partial}
\newcommand{\dd}{\delta}

\newcommand{\new}[1]{#1}

\usepackage{xargs}
\usepackage[colorinlistoftodos,prependcaption,textsize=footnotesize]{todonotes}
\newcommandx{\DM}[2][1=]{\todo[linecolor=orange,backgroundcolor=orange!25,bordercolor=orange,#1]{DM: #2}}
\newcommandx{\KB}[2][1=]{\todo[linecolor=purple,backgroundcolor=purple!25,bordercolor=purple,#1]{KB: #2}}
\newcommandx{\PH}[2][1=]{\todo[linecolor=cyan,backgroundcolor=cyan!25,bordercolor=cyan,#1]{PH: #2}}
\newcommandx{\TODO}[2][1=]{\todo[linecolor=red,backgroundcolor=red!25,bordercolor=red,#1]{TODO: #2}}

\title{Stabilizing complex Langevin for real-time gauge theories with an anisotropic kernel}

\author[a]{Kirill Boguslavski,}
\author[a]{Paul Hotzy,}
\author[a]{David I.~M\"uller}

\affiliation[a]{Institute for Theoretical Physics, TU Wien,\\Wiedner Hauptstraße 8-10, A-1040 Vienna, Austria}

\emailAdd{kirill.boguslavski@tuwien.ac.at}
\emailAdd{paul.hotzy@tuwien.ac.at}
\emailAdd{dmueller@hep.itp.tuwien.ac.at}

\abstract{The complex Langevin (CL) method is a promising approach to overcome the sign problem that occurs in real-time formulations of quantum field theories. Using the Schwinger-Keldysh formalism, we study SU($N_c$) gauge theories with CL. We observe that current stabilization techniques are insufficient to obtain correct results. Therefore, we revise the discretization of the CL equations on complex time contours, find a time reflection symmetric formulation and introduce a novel anisotropic kernel that enables CL simulations on discretized complex time paths. Applying it to SU(2) Yang-Mills theory in 3+1 dimensions, we obtain unprecedentedly stable results that we validate using additional observables and that can be systematically improved. For the first time, we are able to simulate non-Abelian gauge theory on time contours whose real-time extent exceeds its inverse temperature. Thus, our approach may pave the way towards an ab-initio real-time framework of QCD in and out of equilibrium with a potentially large impact on the phenomenology of heavy-ion collisions.
}

\arxivnumber{2212.08602}

\begin{document} 
\maketitle
\flushbottom

\newpage

\section{Introduction}

A full description of the time evolution of quantum field theories remains one of the major unresolved problems in modern physics. For instance, the non-equilibrium evolution of a strongly interacting quark-gluon plasma (QGP) is not fully understood from first principles in Quantum Chromodynamics (QCD). The QGP has existed in the early universe shortly after the Big Bang and can be created artificially in relativistic heavy-ion collisions (the so-called ``Little Bang'') at collider facilities such as the Large Hadron Collider (LHC) and the Relativistic Heavy-Ion Collider (RHIC). Thus, improving our understanding of the dynamical properties of the QGP, especially from an ab-initio perspective, is of high theoretical and phenomenological interest.

Effective models and approaches of QCD have nonetheless led to remarkable phenomenological success in the past few decades. The current theoretical view is that the QGP evolves in a number of separate stages \cite{Berges:2020fwq}.
The early non-equilibrium stage can be described using classical approaches \cite{Schenke:2012wb, Schenke:2012hg, Berges:2013eia, Boguslavski:2019fsb, Ipp:2020igo, Ipp:2021lwz} and kinetic theory \cite{Baier:2000sb,Arnold:2002zm,Kurkela:2018wud} at weak coupling and, at strong coupling, using holography \cite{Chesler:2010bi,Casalderrey-Solana:2013aba}. However, these are limited in applicability and often rely on strong approximations. The subsequent evolution of the QGP is modelled as a relativistic viscous fluid \cite{Gale:2013da, Romatschke:2017ejr}, where the underlying hydrodynamic equations require theoretical input such as viscosity coefficients. Similarly, transport coefficients are important ingredients for modelling jet physics, heavy quarks and quarkonia \cite{Qin:2015srf,Brambilla:2020qwo}. Often such observables are studied using Euclidean lattices along an imaginary time path, where Monte Carlo methods are available \cite{Gattringer:2010zz}, and are analytically continued to real times (see e.g.~\cite{Asakawa:2000tr, Meyer:2011gj, Burnier:2013nla, Altenkort:2022yhb}). 
The analytical continuation is an ill-posed numerical problem because in practice only a limited amount of Euclidean data with finite accuracy is available.
Thus, a general real-time approach for QCD from first principles in and out of equilibrium is highly desirable. 

Essentially, such a formulation requires real-time path integrals in the Schwinger-Keldysh formalism \cite{Schwinger:1960qe,Keldysh:1964ud}. They involve integrands of the form $e^{i S}$, where $S$ is the action of an interacting system. 
In cases where perturbation theory is not applicable, a numerical approach to solving these path integrals becomes necessary.
Unfortunately, due to their highly oscillating nature, these types of integrals are generally intractable with standard numerical methods.
This is known as the infamous sign problem \cite{Gattringer:2016kco} that hinders, or at least complicates direct computations of time-dependent observables. Furthermore, sign problems also occur for systems with complex actions such as QCD at finite density \cite{deForcrand:2009zkb}, which is relevant to heavy-ion collisions and the physics of compact stars.

There are multiple direct approaches to overcome the sign problem (see for instance \cite{Alexandru:2020wrj} for a review). 
Among them, \emph{Complex Langevin} (CL) represents a powerful and very promising method \cite{Aarts:2013uxa,Seiler:2017wvd,Attanasio:2020spv}. 
It is based on a complex extension of stochastic quantization of quantum field theories \cite{Parisi:1980ys, Damgaard:1987rr}, which is formulated as a stochastic differential equation in the fictitious Langevin time. For complex actions, this stochastic process leads to a complexification of the original degrees of freedom of the theory. Provided certain conditions are satisfied \cite{Aarts:2009uq,Nagata:2016vkn}, the process converges to a distribution consistent with the complex path integral weight $e^{iS}$. 
Expectation values of observables can then be computed in terms of Monte Carlo averages by sampling configurations from the converged Langevin process.

In recent years a number of systems have been studied using the CL method such as spin models \cite{Aarts:2011zn, Aarts:2012ft}, Bose gases at finite chemical potential \cite{Aarts:2008wh,Heinen:2022eyh} and scalar fields in real-time \cite{Berges:2005yt,Berges:2006xc, Anzaki:2014hba,Alvestad:2021hsi, Alvestad:2022abf}. One of the major applications of CL has been the investigation of QCD at finite chemical potential \cite{Fromm:2012eb, Sexty:2013ica, Aarts:2017vrv, Seiler:2017wvd,Scherzer:2020kiu,Attanasio:2020spv,Attanasio:2022mjd}.
In contrast, apart from a few pioneering studies \cite{Berges:2006xc, Berges:2007nr, Aarts:2017hqp}, non-Abelian gauge theories in real time have received less attention, since the sign problem associated with real-time paths has been widely believed to be particularly severe. The main problems with CL are numerical instabilities and issues with wrong convergence, i.e.~the stochastic process approaches a wrong stationary solution. These problems are notoriously difficult to solve and a number of methods to reduce them have been proposed in recent years. Often numerical instabilities are related to the discretization of the Langevin equation and can be reduced by adaptive step sizes \cite{Aarts:2009dg} or implicit solvers \cite{Alvestad:2021hsi}. Problems with wrong convergence are mitigated using kernels \cite{Soderberg:1987pd, Okamoto:1988ru, Okano:1991tz, Okano:1992hp, Aarts:2012ft, Alvestad:2022fpl} and, especially in the case of gauge theories, through gauge cooling \cite{Seiler:2012wz, Aarts:2013uxa} and dynamical stabilization \cite{Attanasio:2018rtq}, which have led to substantial progress in the study of the QCD phase diagram.

As an important step towards a genuine real-time formulation of QCD, here we investigate lattice simulations of Yang-Mills theory on real-time contours using the CL approach. Since earlier studies have suffered from severe problems regarding wrong convergence \cite{Berges:2006xc}, we address these issues using modern stabilization methods that are usually applied to QCD at finite density. Although applying these methods leads to remarkable improvements regarding stability on regularized complex time paths \cite{Aarts:2017hqp}, they turn out to be insufficient in the approach towards the actual Schwinger-Keldysh contour. Therefore, we revisit the basic formulation of the CL equations for complex time contours and develop a new anisotropic kernel, which in numerical simulations of SU(2) Yang-Mills theory in 3+1 dimensions leads to unprecedentedly stable results with correct convergence. In particular, we find that the stabilizing effect of the kernel leads to metastable regions in Langevin time, which can be extended systematically at the expense of increased lattice sizes. The metastable regions are long-lived enough that, for the first time, we are able to obtain correct results on complex time paths whose real-time extent exceeds the inverse temperature. Our new CL equations thus mark a substantial conceptual progress that may allow us to calculate real-time observables in QCD from first principles with important applications in heavy-ion physics.

This paper is organized as follows. We first revisit the formulation of CL on complex time paths for a simple quantum mechanical toy model in \se\ref{sec:toy_model}. To resolve ambiguities in the CL equation, our formulation is based on parameterizing the complex time contour. We extend our approach to lattice gauge theory in \se\ref{sec:revCL_rtYM}. Motivated by the continuum limit of the Schwinger-Keldysh contour, we exploit the kernel freedom of CL to arrive at new CL update equations using an anisotropic kernel. In \se\ref{sec:results} we demonstrate and validate that this approach leads to remarkably stable results when applied to SU(2) Yang-Mills theory in 3+1 dimensions. We conclude this work with a summary of our findings and an outlook for our future studies in \se\ref{sec:conclusion}.
\new{Additional details on CL formulations with contour parametrization, their discretized equations and useful stabilization techniques can be found in the Appendices.}

\section{Real-time complex Langevin for a simple system} 
\label{sec:toy_model}

    \new{In this section we discuss the CL method, its application to a quantum mechanical model on a complex time contour and resulting ambiguities from a naive formulation. We resolve these issues in}~\se\ref{sec:param_contour} \new{where we also employ the kernel freedom to reproduce the Minkowksi and the Euclidean cases. An unambiguous discretized CL equation is introduced in}~\app \ref{app:disc_drift}.

    \subsection{Introducing the Complex Langevin method}
    
    The dynamics of quantum mechanical systems can be described using Feynman's path integral approach. In its real-time path integral formulation, one typically encounters oscillatory integrals for expectation values of the form
    \begin{align} \label{eq:osc_integral}
        \langle \mathscr{O} \rangle = \frac{1}{Z} \int d x\, \mathscr{O}(x) \exp \left[ i S(x) \right],\qquad Z = \int d x\, \exp \left[ i S(x) \right],
    \end{align}
    for an observable $\mathscr{O}(x)$ where $S(x)$ is the classical action of the theory. While such integrals can be solved analytically for simple systems, the numerical computation of expectation values for more complicated systems is typically not feasible with standard methods such as Monte Carlo integration or numerical integration methods based on discretization. The former is not applicable since $\exp\left[ i S(x) \right]$ can not be interpreted as a probability density, and the latter fails due to the highly oscillatory nature of the integrand, also known as the sign problem. The complex Langevin method is an approach to overcome this issue. It introduces an artificial auxiliary time coordinate $\theta$ known as the Langevin time and uses complexified degrees of freedom, i.e., $x \mapsto z(\theta) \in \mathbb C$. The evolution along $\theta$ is described by a complex stochastic process known as the CL equation and is given by
    \begin{align} \label{eq:cle_simple}
        \frac{dz(\theta)}{d\theta} &= i\frac{dS}{dz} + \eta(\theta),
    \end{align}
    where $dS/dz$ is called the drift term. The action $S$ for complex arguments $z$ is understood to be the analytic continuation of $S(x)$. The real-valued noise term $\eta(\theta)$ is Gaussian distributed and satisfies
    \begin{align} \label{eq:cle_simple_noise}
        \langle \eta(\theta) \rangle = 0,\qquad \langle \eta(\theta) \eta(\theta^\prime) \rangle = 2 \delta(\theta-\theta^\prime).
    \end{align}
    If the stochastic process described by the CL equation converges (see \cite{Aarts:2009uq,Nagata:2016vkn} for discussions on the assumptions), the distribution of the stochastic variable $z$ approaches the stationary solution of the corresponding complex Fokker-Planck equation. In this case, the calculation of expectation values in \eq\eqref{eq:osc_integral} can be equivalently replaced by sampling the stochastic process at large Langevin times $\theta$ 
    \begin{align} \label{eq:sample}
        \langle \mathscr{O} \rangle \approx \lim\limits_{\theta_0\rightarrow\infty} \frac{1}{T}\int_{\theta_0}^{\theta_0+T}  d \theta\, \mathscr{O}[z(\theta)].
    \end{align}
    In this way, the CL method can circumvent the sign problem. However, additional ambiguities are encountered when we consider the real-time path integral formulation. In \new{the following} we discuss and resolve them for a simple quantum mechanical system before continuing with more complicated gauge field theories in the following sections.

    \subsection{Simple model on a complex time contour}
    \label{sec:simple_model}
\begin{figure} [t]
            \centering
            \begin{subfigure}[t]{.45\textwidth}
                \centering\includegraphics[width=1\linewidth]{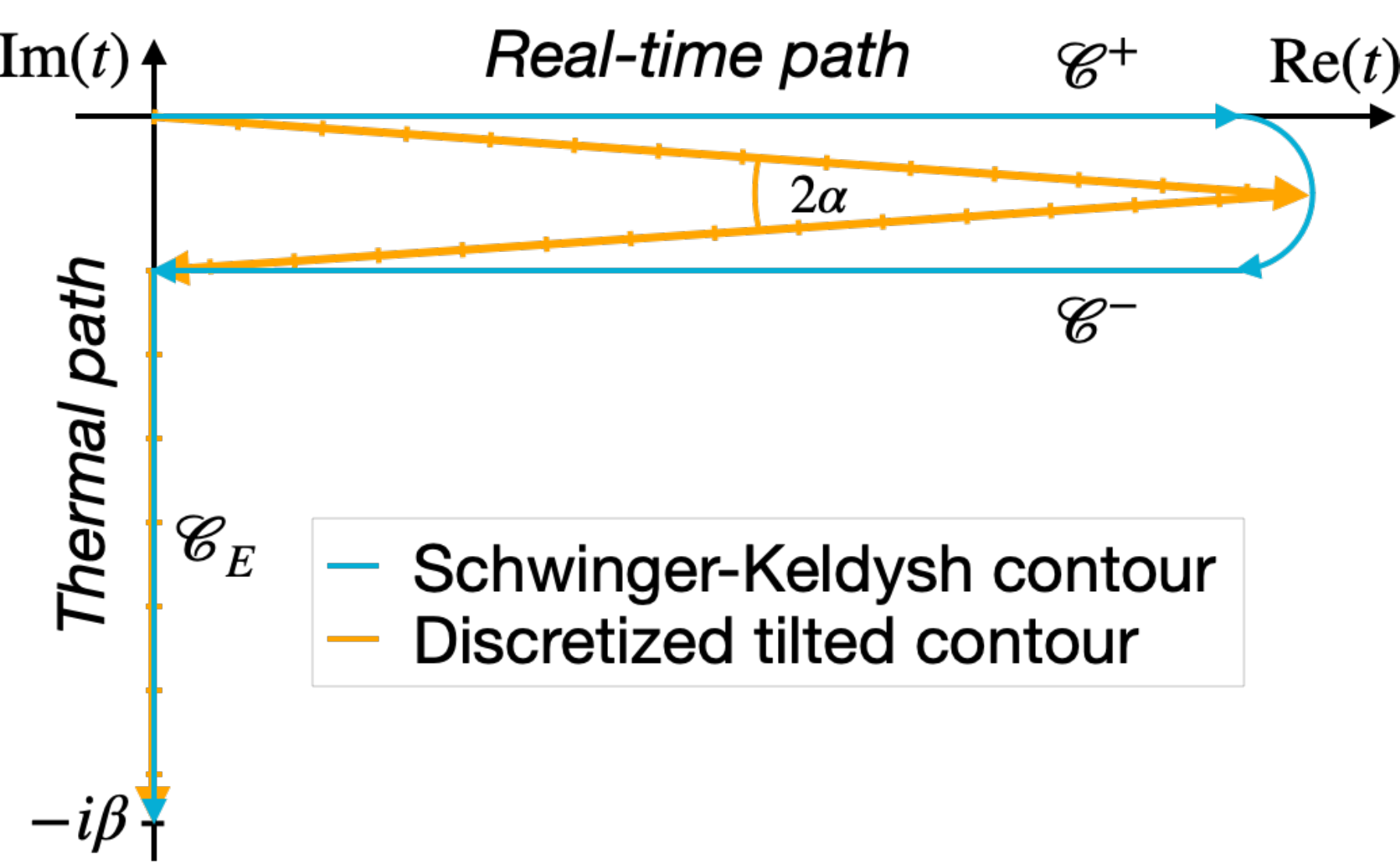}
                \caption{\label{fig:disc_vs_cont_contour}}
            \end{subfigure}
            \hfill
            \begin{subfigure}[t]{.45\textwidth}
                \centering\includegraphics[width=1.\linewidth]{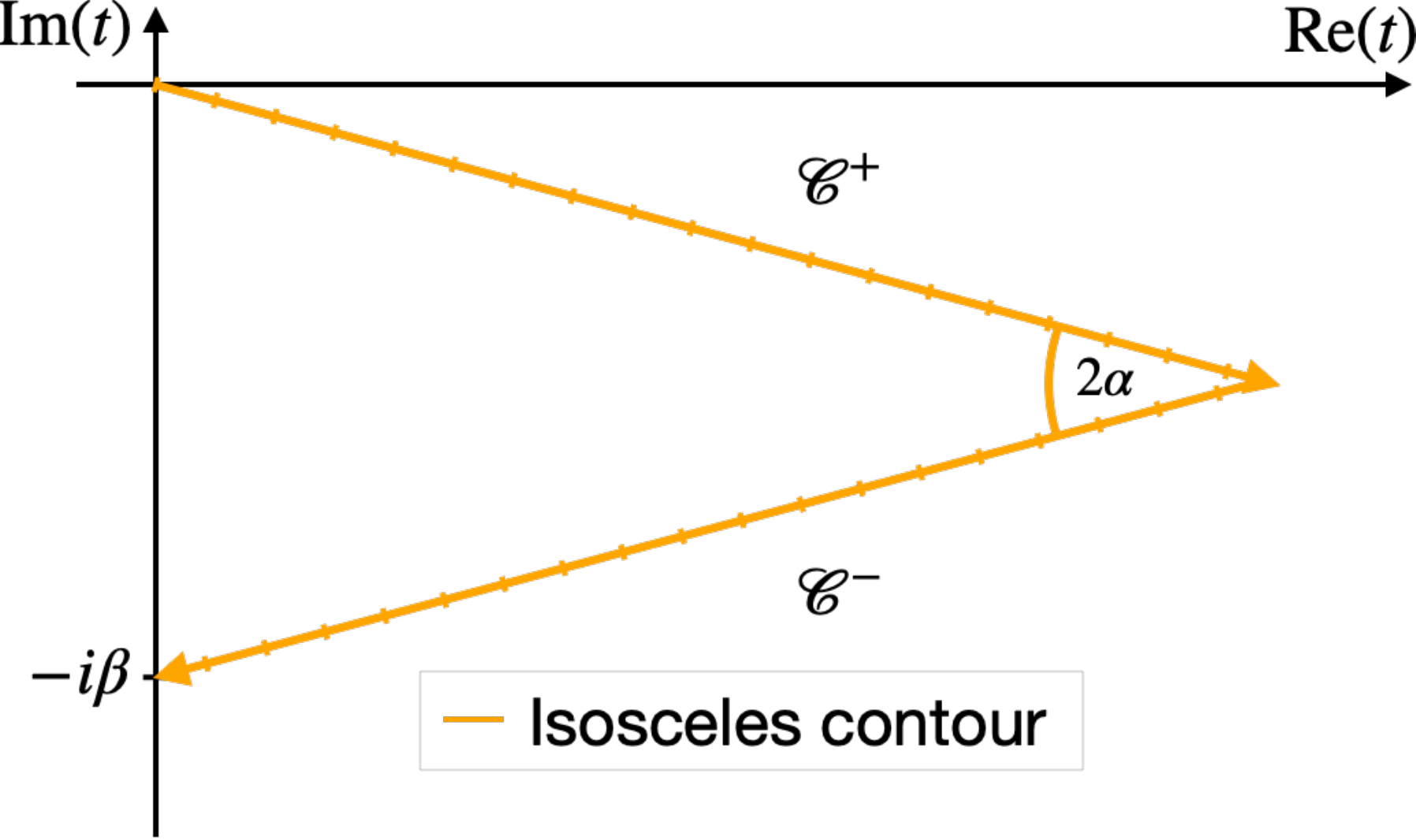}
                \caption{\label{fig:isosceles_contour}}
            \end{subfigure}
            \caption{(a) The Schwinger-Keldysh time contour and a discretized tilted time-contour are shown as blue and orange lines, respectively. The real-time part consists of the two branches $\mathscr{C}^+$ and $\mathscr{C}^-$ while the thermal Euclidean path is denoted by $\mathscr{C}_E$. (b) Isosceles time contour with tilt angle $\alpha$. 
            \label{fig:contour}}
        \end{figure}

        As a concrete example, we consider a quantum mechanical model in thermal equilibrium with dependence on real time within the Schwinger-Keldysh formalism. The action is given by
        \begin{align} \label{eq:mechnical_action}
            S\left[x(t)\right] = \intop_\mathscr{C} dt \, \left[ \frac{1}{2} \left(\frac{dx}{dt}\right)^2 - V(x(t)) \right],
        \end{align}
        where $x(t)$ is the trajectory of a particle with mass $m=1$ in a potential $V$. The complex contour path $\mathscr{C}$ that is integrated over in \eq\eqref{eq:mechnical_action} denotes the Schwinger-Keldysh contour \cite{Schwinger:1960qe,Keldysh:1964ud} and is assumed to be continuous, starting at $t=0$ and ending at $t=-i\beta$ without crossings. Here, $\beta$ denotes the inverse temperature of the system. A sketch of the contour is visualized as the blue curve shown in Fig.\ \ref{fig:disc_vs_cont_contour}.
        The Schwinger-Keldysh formalism allows us to calculate expectation values via the path integral
        \begin{align} \label{eq:mechanical_expectation_value}
            \langle \mathscr{O}[x] \rangle = \frac{1}{Z}\int \mathcal{D}x_E \, e^{-S_E[x_E]} \int \mathcal{D}x_+\, \mathcal{D}x_-\, e^{i S[x_+,x_-]} \, \mathscr{O}(x)\,.
        \end{align}
        Here, $x_+$ and $x_-$ denote the trajectories on the forward and backward real-time paths $\mathscr{C}^+$ and $\mathscr{C}^-$, respectively, while $x_E$ is defined on the Euclidean (purely imaginary) part of the contour $\mathscr{C}_E$. 
        All times on $\mathscr{C}^-$ are considered later than on $\mathscr{C}^+$ and both real-time branches are separated infinitesimally along the imaginary time direction.
        At the start and endpoints of the contour, $x(t)$ satisfies periodic boundary conditions
        \begin{align}
            x (t{=}0) = x (t{=}-i\beta).
        \end{align}
        In our model \eqref{eq:mechnical_action}, the real-time paths in \eq\eqref{eq:mechanical_expectation_value} lead to highly oscillatory integrals similar to \eq\eqref{eq:osc_integral}.
        
    \subsection{Ambiguities for complex time contours} \label{sec:ambiguities}
        
        Our goal is here to formulate the CL equation for the model in \se\ref{sec:simple_model} by complexifying the trajectory of the particle and introducing the auxiliary Langevin time $\theta$
        \begin{align}
            x(t) \rightarrow z(\theta, t).
        \end{align}
        Compared to the stochastic process described by \eq \eqref{eq:cle_simple}, we have introduced an additional dependence on the physical time $t \in \mathbb C$. A straightforward but naive generalization of \eq \eqref{eq:cle_simple} to the Schwinger-Keldysh time contour reads
        \begin{align} \label{eq:cle_SK_naive}
            \begin{split}
            \frac{d z(\theta, t)}{d \theta} &= i \left. \frac{\delta S}{\delta z(t)} \right\vert_{\theta} + \eta(\theta, t), \\
            \langle \eta(\theta, t) \rangle &= 0,\\
            \langle \eta(\theta, t) \eta(\theta', t') \rangle &= 2 \delta(\theta - \theta') \delta(t - t'),
            \end{split}
        \end{align}
        where $\vert_{\theta}$ is a shorthand notation for the replacement $z(t) \rightarrow z(\theta, t)$ after differentiation, i.e., denoting that the drift term is evaluated for fields at Langevin time $\theta$. 
        The issue with this formulation is that the noise correlator at $t-t'$ for some points $t$, $t'$ on the complex contour is ambiguous because the Dirac delta distribution is usually defined for real-valued arguments. Thus, the noise correlator must be adapted in order to obtain an unambiguous CL equation for complex time contours.
        
        However, there are at least two special cases, where this ambiguity does not arise: the Minkowski and the Euclidean time contours.
        The CL equation for Minkowski time $t \in \mathbb R$ generalizes \eq \eqref{eq:cle_simple} in a natural way
        \begin{align} \label{eq:cle_mink}
            \begin{split}
            \frac{d z(\theta, t)}{d \theta}  &= i \left. \frac{\delta S_M}{\delta z(t)} \right\vert_{\theta} + \eta(\theta, t), \\
            \langle \eta(\theta, t) \rangle &= 0, \\
            \langle \eta(\theta, t) \eta(\theta', t') \rangle &= 2 \delta(\theta - \theta') \delta(t - t'),
            \end{split}
        \end{align}
        with the Minkowski action
        \begin{align} 
            S_M\left[x(t)\right] = \intop_{-\infty}^{+\infty} dt \, \left[ \frac{1}{2} \left(\frac{dx}{dt}\right)^2 - V(x(t)) \right].
        \end{align}
        On the other hand, for a Euclidean time contour $t= -i\tau$ with $\tau \in [0, \beta]$ we may write
        \begin{align} \label{eq:cle_eucl}
            \begin{split}
            \frac{d z(\theta, \tau)}{d \theta}  &= - \left. \frac{\delta S_E}{\delta z(\tau)} \right\vert_{\theta} + \eta(\theta, \tau), \\
            \langle \eta(\theta, \tau) \rangle &= 0, \\
            \langle \eta(\theta, \tau) \eta(\theta', \tau') \rangle &= 2 \delta(\theta - \theta') \delta(\tau - \tau')\,,
            \end{split}
        \end{align}
        where $S_E$ denotes the Euclidean action
        \begin{align}
            S_E[x(\tau)] = \intop_0^\beta d\tau \, \left[ \frac{1}{2} \left(\frac{dx}{d\tau}\right)^2 + V(x(\tau)) \right].
        \end{align}
        Since there appear no complex terms in the Euclidean case, the stochastic process stays real and the CL equation reduces to the real Langevin equation.
        
        Evidently, both systems can be formulated without ambiguities, because the arguments in the noise correlator are always real-valued. Thus, our goal is to find a consistent formulation of CL which not only correctly reproduces both Minkowski and Euclidean contours as limiting cases, but is well-defined for any parametrizable complex time contour such as the Schwinger-Keldysh contour.

    \subsection{Parametrizing the time contour \& kernel freedom} \label{sec:param_contour}
        The ambiguities from the Dirac delta distribution in the noise correlator can be resolved by choosing an explicit bijective parametrization of the time contour $\mathscr{C}$ with $\lambda \mapsto t(\lambda)$ on
        \begin{align}
            t: [a, b] \mapsto \mathbb{C}, \qquad t(a) = t_0, \quad t(b) = t_1,
        \end{align}
        where $t_0$ and $t_1$ are the start and end points. 
        An unambiguous CL equation, written in terms of the curve parametrization, is given by
        \begin{align} \label{eq:cle_lambda_original}
            \frac{d z(\theta, \lambda)}{d\theta} = i \left. \frac{\delta S}{\delta z(\lambda)} \right\vert_{\theta} + \eta(\theta, \lambda),
        \end{align}
        where the noise satisfies
        \begin{align} \label{eq:lambda_noise}
        \begin{split}
            \langle \eta(\theta, \lambda) \rangle &= 0, \\
            \langle \eta(\theta, \lambda) \eta(\theta', \lambda') \rangle &= 2 \delta(\theta - \theta') \delta(\lambda - \lambda').
        \end{split}
        \end{align}
        Here, the noise correlator is well-defined since the curve parameters $\lambda$, $\lambda'$ are real-valued. In the following we demonstrate that the parameterized CL equation correctly reproduces the Minkowksi and the Euclidean case and is invariant under reparameterization.
        
        The action in terms of an integral along the contour parameter $\lambda$ reads
        \begin{align} \label{eq:lambda_action}
            S[x(\lambda)] = \intop^{a}_{b} d\lambda \, \frac{d t}{d\lambda} \bigg( \frac{1}{2} \left(\frac{dx}{dt}\right)^2 - V(x(t)) \bigg),
        \end{align}
        where we have kept derivatives of the trajectories $x$ in terms of $t$ instead of the curve parameter $\lambda$ for brevity. Computing the variation
        \begin{align} \label{eq:lambda_variation}
            \delta S[x(\lambda), \delta x(\lambda)] = \intop^{b}_{a} d\lambda \, \frac{\delta S}{\delta x(\lambda)} \delta x(\lambda) = \intop^{b}_{a} d\lambda \, \frac{dt}{d\lambda} \left(
            - \frac{d^2x}{dt^2} - V'\left(x(t)\right) \right) \delta x(\lambda),
        \end{align}
        yields the drift term
        \begin{align} \label{eq:lambda_drift}
            i  \frac{\delta S}{\delta x(\lambda)} = - i \frac{dt}{d\lambda} \left(
             \frac{d^2x}{dt^2} + V'(x(t)) \right).
        \end{align} 
        Note that the functional derivatives with respect to $x(\lambda)$ and $x(t)$ are related by the derivative of the curve parametrization $dt/d\lambda$
        \begin{align}
            \frac{\delta S}{\delta x(\lambda)} = \frac{dt}{d\lambda} \frac{\delta S}{\delta x(t)}.
            \label{eq:dSdxl_dSdxt}
        \end{align}
        The CL equation for the parametrized time contour in terms of the usual drift term thus reads
        \begin{align}
            \frac{d z(\theta, t)}{d\theta} = i \frac{dt}{d\lambda} \left. \frac{\delta S}{\delta z(t)} \right\vert_{\theta} + \eta(\theta, \lambda).
        \end{align}
        In the above, the variable $z(\theta, t)$ is understood to mean $z(\theta, \lambda(t))$.
At this point we may ask whether the CL equation formulated for general complex time contours indeed reduces to the Minkowksi and Euclidean cases.
        
Starting with the Minkowski case, we choose some \new{real-valued parametrization \mbox{$t: \mathbb{R} \to \mathbb{R}$}}. Since the parametrization must be bijective, we require $dt/d\lambda > 0$ along the contour. Equation \eqref{eq:dSdxl_dSdxt} relates the drift term of the parametrized action to the usual Minkowski action. Similarly, we may relate the noise correlator of \eq\eqref{eq:lambda_noise} in terms of $\lambda$ to the expression for Minkowski time $t$ in \eq\eqref{eq:cle_mink} and obtain
        \begin{align}
            \langle \eta(\theta, \lambda) \eta(\theta', \lambda') \rangle = 2 \frac{dt}{d\lambda} \delta(\theta - \theta') \delta(t - t') = \frac{dt}{d\lambda} \langle \eta(\theta, t) \eta(\theta', t') \rangle ,
        \end{align}
        where $t = t(\lambda)$ and $t'=t(\lambda')$. The above relation between the correlators suggests the following transformation between $\eta(\theta, \lambda)$ and $\eta(\theta, t)$:
        \begin{align}
            \eta(\theta, \lambda) = \sqrt{\frac{dt}{d\lambda}}\, \eta(\theta, t).
        \end{align}
        Inserting both the drift term and the transformed noise into the CL equation in \eq\eqref{eq:cle_lambda} yields
        \begin{align} \label{eq:cle_lambda}
            \frac{d z(\theta, t)}{d\theta} = 
            i \frac{d t}{d\lambda} \left. \frac{\delta S_M}{\delta z(t)} \right\vert_{\theta} + \sqrt{\frac{dt}{d\lambda}} \eta(\theta, t).
        \end{align}
        
        The above CL equation includes additional $dt/d\lambda$ factors in front of the drift and noise terms when compared to the original stochastic process defined in \eq\eqref{eq:cle_mink}. However, different CL equations can be grouped into equivalence classes, where two equations are equivalent if the stationary solution in the limit $\theta \rightarrow \infty$ is described by the same probability distribution. The correspondence between Langevin equations and their stationary Fokker-Planck equation is only unique up to the so-called kernel freedom (see, e.g., chapter 4 of \cite{Namiki:1993fd}). More precisely, all Langevin equations of the following form approach the same stationary solution of the corresponding Fokker-Planck equations:
        \begin{align} \label{eq:cle_kernel}
            \frac{d z(\theta, t)}{d \theta} = i \int dt' \, \Gamma(t,t') \left. \frac{\delta S}{\delta z(t')} \right\vert_{\theta} + \int dt' \, \tilde \Gamma(t,t') \eta(\theta, t'), 
        \end{align}
        where $\Gamma(t,t')$ denotes a field-independent kernel which is required to be factorizable
        \begin{align}
            \Gamma(t,t') = \int dt'' \, \tilde \Gamma(t, t'') \tilde \Gamma(t', t'')\,.
        \end{align}
        Comparing \eq\eqref{eq:cle_lambda} with the original formulation in \eq\eqref{eq:cle_mink}, we notice that the additional factors correspond to a field-independent kernel 
        \begin{align}
            \Gamma(t, t') = \delta(t-t')\, \frac{d t}{d\lambda}.
        \end{align}
        Hence, the formulation of the CL equation in terms of a parametrization of the Minkowksi time in \eq\eqref{eq:cle_lambda} is equivalent to the CL equation \eq\eqref{eq:cle_mink} with physical time $t$.
        
        An analogous argument can be made for the Euclidean case. Here we choose $t= - i\tau(\lambda)$ where $\tau(\lambda) \in [0, \beta]$ parameterizes the imaginary part of the time contour with $d\tau/d\lambda > 0$. The drift term is
        \begin{align}
            i\frac{\delta S}{\delta x(\lambda)} = i\frac{dt}{d\lambda} \frac{\delta S}{\delta x(t)} =  \frac{d\tau}{d\lambda} \frac{\delta S}{\delta x(t)} = - \frac{d\tau}{d\lambda} \frac{\delta S_E}{\delta x(\tau)}, 
        \end{align}
        where we have used \eq\eqref{eq:dSdxl_dSdxt} and $S = -i S_E$ due to $dt = - i d\tau$. Similarly, the noise correlator reads
        \begin{align}
            \langle \eta(\theta, \lambda) \eta(\theta', \lambda') \rangle = 2 \frac{d\tau}{d\lambda} \delta(\theta - \theta') \delta(\tau - \tau') = \frac{d\tau}{d\lambda} \langle \eta(\theta, \tau) \eta(\theta', \tau') \rangle.
        \end{align}
        The parametrized Euclidean CL equation can thus be written as
        \begin{align}            
            \frac{d z(\theta, \tau)}{d\theta} = 
            - \frac{d \tau}{d\lambda} \left. \frac{\delta S_E}{\delta z(\tau)} \right\vert_{\theta} + \sqrt{\frac{d\tau}{d\lambda}} \,\eta(\theta, \tau).
        \end{align}
        The above equation is equivalent to the original Euclidean process in \eq\eqref{eq:cle_eucl} up to a kernel
        \begin{align}
            \Gamma(\tau, \tau') = \delta(\tau-\tau')\, \frac{d \tau}{d\lambda}.
        \end{align}
        
        Having shown that the parameterized CL equation in \eq\eqref{eq:cle_lambda_original} correctly reproduces both the Minkowski and the Euclidean case and, by virtue of the parameterization, has an unambiguous noise correlator for general complex time contours $t(\lambda) \in \mathbb C$, we posit that it is the right approach to formulate the CL method for the Schwinger-Keldysh contour.
        
        Moreover, we emphasize that one can similarly show that CL equations with different parametrizations are related by a kernel and thereby yield the same stationary solution in the case of convergence. 
        Starting from \eq\eqref{eq:cle_lambda_original} and introducing a reparametrization $\lambda \mapsto \xi(\lambda)$, $t(\lambda) \to t(\xi) = t(\xi(\lambda))$ with $d\xi/d\lambda > 0$, we can perform the same steps as before to find the reparameterized CL equation
        \begin{align}
            \frac{d z(\theta, \xi)}{d\theta} = i \frac{d\xi}{d\lambda} \left. \frac{\delta S}{\delta z(\xi)} \right\vert_{\theta} + \sqrt{\frac{d\xi}{d\lambda} }\,\eta(\theta, \xi),
        \end{align}
        with
        \begin{align}
            \langle \eta(\theta, \xi) \eta(\theta', \xi') \rangle &= 2 \delta(\theta - \theta') \delta(\xi - \xi'),
        \end{align}
        which is kernel-equivalent to the CL equation in terms of $\xi$
        \begin{align}
            \frac{d z(\theta, \xi)}{d\theta} = i \left. \frac{\delta S}{\delta z(\xi)} \right\vert_{\theta} + \eta(\theta, \xi).
        \end{align}
        Hence, the parameterized CL equation does not depend on the specific parametrization of the time contour provided that it converges to the stationary solution.

    In order to numerically simulate the stochastic process defined by the CL equation, we have to discretize the trajectory $x(t)$ along the time contour. We choose a time-reversal symmetric discretization of the action \eq \eqref{eq:lambda_action}
        \begin{align} \label{eq:lambda_action_discr}
            S_{\mathrm{latt}} [x] = \sum_{k=1}^{N_t} (\lambda_{k} - \lambda_{k-1}) \frac{t_{k} - t_{k-1}}{\lambda_{k} - \lambda_{k-1}} \left( \frac{1}{2} \left( \frac{x_{k} - x_{k-1}}{t_{k} - t_{k-1}} \right)^2 - \frac{1}{2} \left(V(x_{k}) + V(x_{k-1}) \right)
            \right),
        \end{align}
        where $t_k = t(\lambda_k) \in \mathbb{C}$, $x_k = x(t_k)$ for $k=0,\dots,N_t$, and one identifies $x_{0} \equiv x_{N_t}$ due to the periodic boundary conditions. With the time steps $a_{t,k} = t_{k+1} - t_k$, and parameterizing the time path by its arc length with the step $a_{\lambda,k} = \lambda_{k+1} - \lambda_k = \vert a_{t,k} \vert$, we arrive in \app \ref{app:disc_drift} at the discrete CL update step in the Euler-Maruyama scheme 
        \begin{align} \label{eq:lambda_cle_disc}
            z_k(\theta + \epsilon) = z_k(\theta) + 
            i \frac{\epsilon}{\frac{1}{2}(|a_{t,k}| + |a_{t,k-1}|)} \left. \frac{\partial S_\mathrm{latt}}{\partial z_k} \right\vert_{\theta}  +
            \sqrt{\frac{\epsilon}{\frac{1}{2}(|a_{t,k}| + |a_{t,k-1}|}}\, \tilde \eta_k(\theta),
        \end{align}
        where $\epsilon$ denotes the Langevin time step and the discrete noise satisfies
        \begin{align}
            \langle \tilde \eta_{k'}(\theta') \tilde \eta_k(\theta) \rangle = 2 \delta_{\theta \theta'} \delta_{kk'}.
        \end{align}
        This is consistent with the update step used in the literature \cite{Alvestad:2021hsi}, where CL was applied to an anharmonic oscillator on a Schwinger-Keldysh contour.

        As mentioned in \re\cite{Alvestad:2021hsi}, the \eq\eqref{eq:lambda_cle_disc} is related to a discrete CL equation without a contour parametrization by interpreting the additional factors as a rescaling of the Langevin time step
        \begin{align}
            \epsilon \quad \mapsto \quad \tilde \epsilon = \frac{1}{\frac{1}{2}(|a_{t,k}| + |a_{t,k-1}|)} \,\epsilon.
        \end{align}
        This can be understood as a discretized version of the kernel freedom discussed above.
    Consequently, both equations will converge to the same stationary solution in the limit of a small Langevin time step $\epsilon \to 0$. However, we want to emphasize that the instabilities CL suffers from might be mitigated or even aggravated by some kerneled CL equations, as the equivalence only holds for $\epsilon \to 0$ at large Langevin times $\theta \to \infty$. In \se\ref{sec:new_method} we will exploit the kernel freedom for gauge theories in a similar way to effectively stabilize the CL equation for real-time Yang-Mills theories.

\section{Revisiting the CL method for real-time Yang-Mills theory}
\label{sec:revCL_rtYM}

    In this section we formulate the CL method for non-Abelian gauge fields on a complex Schwinger-Keldysh time contour $\mathscr{C}$, depicted as the blue curve in \fig \ref{fig:disc_vs_cont_contour}. The Yang-Mills action is given by
    \begin{align} \label{eq:ym_action}
        S_\mathrm{YM} = - \frac{1}{4} \int_{\mathscr C} d^{d+1}x F^{\mu \nu}_a F_{\mu \nu}^a,
    \end{align}
    with the non-Abelian field strength tensor
    \begin{align}
        F_{\mu \nu}^a = \partial_\mu A^a_\nu - \partial_\nu A^a_\mu - g f^{abc} A^b_\mu A^c_\mu, 
    \end{align}
    and totally antisymmetric structure constants $f^{abc}$. The coupling constant is given by $g$. Throughout this work we assume implicit summation over Lorentz indices $\mu,\nu = 0,1,\dots, d$ and color indices $a, b = 1,\dots,N_c^2-1$. 

    Employing the Schwinger-Keldysh formalism, our goal is to compute expectation values
    \begin{align} \label{eq:pi_gauge}
        \langle \mathscr{O}[A] \rangle = \frac{1}{Z}\int \mathcal{D}A_E \, e^{-S_E[A_E]} \int \mathcal{D}A_+\, \mathcal{D}A_-\, e^{i S[A_+,A_-]} \, \mathscr{O}(x),
    \end{align}
    using the CL method, for gauge fields satisfying periodic boundary conditions
    \begin{align}
        A^a_\mu (t=0) = A^a_\mu (t=-i\beta).
    \end{align}

    In the following, we first state the CL equation for Yang-Mills theory on complex time contours in \se\ref{sec:CL_YM_cont}, reformulate it by parametrizing the time contour in \se\ref{sec:ym_contour_param}, discretize the resulting CL equation in \se\ref{sec:lattice} and compare our strategy to previous approaches in \se\ref{sec:early_ym_cle}. To mitigate instabilities, we introduce a new anisotropic CL kernel in \se\ref{sec:new_method} and explain how observables on the Schwinger-Keldysh time contour can be obtained systematically. 
    The main result of this section are the CL equations in \eqs\eqref{eq:new_cle_temp} and \eqref{eq:new_cle_spat} that we will use in our numerical simulations in \se\ref{sec:results}.
        
    \subsection{Complex Langevin for Yang-Mills theories on complex time contours}
    \label{sec:CL_YM_cont}
        Applying the CL method to $\mathrm{SU}(N_c)$ gauge theories requires the complexification of the corresponding Lie algebra $\mathfrak{su}(N_c) \to \mathfrak{sl}(N_c, \mathbb{C})$.
        The CL equation for Yang-Mills theory on complex time contours can be naively formulated as
        \begin{align} \label{eq:ym_cle_naive}
        \begin{split}
            \frac{\partial A^a_\mu(\theta, t, \mx)}{\partial \theta} &= i \left. \frac{\delta S_\mathrm{YM}}{\delta A^a_\mu(t, \mx)}\right\vert_{\theta} + \eta^a_\mu(\theta, t, \mx), \\
            \langle \eta^a_\mu(\theta, t, \mx) \rangle &= 0, \\
            \langle \eta^a_\mu(\theta, t, \mx) \eta^b_\nu(\theta', t', \mx') \rangle &= 2 \delta(\theta - \theta') \delta(t-t') \delta^{(d)}(\mx-\mx') \delta^{ab} \delta_{\mu\nu},
        \end{split}
        \end{align}
        where $\mx, \mx' \in \mathbb R^{(d)}$ denote spatial coordinates and the subscript $\vert_{\theta}$ implies that the drift term is evaluated for fields at Langevin time $\theta$. 
        The direction and color dependent noise term $\eta^a_\mu(\theta, x)$ is governed by a Gaussian distribution in each degree of freedom. We face the same issues with complex times $t,t'\in \mathbb C$ as in \se\ref{sec:ambiguities} because of the Dirac distribution in the noise correlator. Hence, the equation above is only consistent for Minkowski time contours where we have $t \in \mathbb{R}$ and replace $S_{\mathrm{YM}}$ with the action in Minkowski time $S_M$. The Minkowski CL equations read
        \begin{align} \label{eq:ym_cle_mink}
        \begin{split}
            \frac{\partial A^a_\mu(\theta, t, \mx)}{\partial \theta} &= i \left. \frac{\delta S_\mathrm{M}}{\delta A^a_\mu(t, \mx)}\right\vert_{\theta} + \eta^a_\mu(\theta, t, \mx), \\
            \langle \eta^a_\mu(\theta, t, \mx) \rangle &= 0, \\
            \langle \eta^a_\mu(\theta, t, \mx) \eta^b_\nu(\theta', t', \mx') \rangle &= 2 \delta(\theta - \theta') \delta(t-t') \delta^{(d)}(\mx-\mx') \delta^{ab} \delta_{\mu\nu},
        \end{split}
        \end{align}
        with the Minkowski action
        \begin{align} 
        S_\mathrm{M} = - \frac{1}{4} \int_{-\infty}^{\infty} dt \int d^dx\, F^{\mu \nu}_a F_{\mu \nu}^a.
    \end{align}
        Similarly, there is a natural way to write down the evolution equation in Euclidean time $t=-i\tau$ with $\tau \in [0,\beta]$, which reads
        \begin{align} \label{eq:ym_cle_eucl}
        \begin{split}
            \frac{\partial A^a_\mu(\theta, \tau, \mx)}{\partial \theta} &= - \left. \frac{\delta S_\mathrm{E}}{\delta A^a_\mu(\tau, \mx)}\right\vert_{\theta} + \eta^a_\mu(\theta, \tau, \mx) \\
            \langle \eta^a_\mu(\theta, \tau, \mx) \rangle &= 0, \\
            \langle \eta^a_\mu(\theta, \tau, \mx) \eta^b_\nu(\theta', \tau', \mx') \rangle &= 2 \delta(\theta - \theta') \delta(\tau-\tau') \delta^{(d)}(\mx-\mx') \delta^{ab} \delta_{\mu\nu},
        \end{split}
        \end{align}
        where $S_E$ denotes the Yang-Mills action in Euclidean time
        \begin{align} 
        S_\mathrm{E} = \frac{1}{4} \int_{0}^{\beta} d\tau \int d^dx\, F^{mn}_a F_{mn}^a.
        \end{align}
        Here summation over the Euclidean indices $m, n = 1, \dots, d+1$ with $A_0 = i A_{d+1}$ is implied.
        
        Having resolved the ambiguities of the Dirac distribution for the toy model in \se\ref{sec:toy_model}, we follow the same strategy to arrive at a consistent formulation for non-Abelian gauge theories on a general complex time contour. As before,  we demand that the stationary solutions for Minkowksi and Euclidean time contours are retained. In the following section this is achieved in analogy to \se\ref{sec:param_contour} by introducing a contour parametrization. 
        However, we will additionally need to take into account that the coordinate transformation $\lambda \rightarrow t(\lambda)$ also induces a coordinate change for the gauge fields $A_\mu$.

    \subsection{Parameterizing the time contour} 
    \label{sec:ym_contour_param}
        
        In the spirit of \se\ref{sec:param_contour}, we parameterize the time contour $\lambda \mapsto t(\lambda)$ by a real-valued curve parameter $\lambda$. The CL equations for gauge fields, written in terms of the contour parameter, read
        \begin{align} \label{eq:lambda_ym_cle}
        \begin{split}
            \p_\theta A^a_\mu(\theta, \lambda, \mathbf x) &= i \left. \frac{\dd S_\mathrm{YM}}{\dd A^a_\mu(\lambda, \mathbf x)} \right\vert_{\theta} + \eta^a_\mu(\theta, \lambda, \mathbf  x), \\
            \langle \eta^a_\mu(\theta, \lambda, \mathbf  x) \rangle &= 0, \\
            \langle \eta^a_\mu(\theta, \lambda, \mathbf x) \eta^b_\nu(\theta', \lambda', \mathbf x') \rangle &= 2 \delta_{\mu\nu} \delta^{ab} \delta(\theta - \theta') \delta(\lambda - \lambda') \delta^{(3)}(\mathbf x- \mathbf x')\,,
        \end{split}
        \end{align}
        where $\mu, \nu \in \{ \lambda, x, y, z\}$ and we have set the number of spatial dimensions to $d=3$ for definiteness. Different from the naive formulation in \eq\eqref{eq:ym_cle_naive}, the contour parameter formulation in \eq\eqref{eq:lambda_ym_cle} involves an unambiguous Dirac delta distribution $\delta(\lambda - \lambda')$. 
        Note that the evolution equation for $\mu=\lambda$ is written in terms of the $\lambda$-component $A^a_\lambda(\lambda, \mx)$ instead of the temporal component $A^a_t(t, \mx)$. 
        As explained in \app \ref{app:ym_contour_param}, we can write \eq\eqref{eq:lambda_ym_cle} in terms of the functional derivative with respect to fields in physical time
        \begin{align} \label{eq:cle_ym_contour_hybrid}
        \begin{split}
            \p_\theta A^a_t(\theta, t, \mx) &= i \frac{d\lambda}{dt}\left.\frac{\dd S_\mathrm{YM}}{\dd A^a_t(t, \mx)}\right\vert_{\theta} + \frac{d\lambda}{dt}\, \eta^a_\lambda(\theta, \lambda, \mx), \\
            \p_\theta A^a_i(\theta, t, \mx) &= i \frac{dt}{d\lambda}\left.\frac{\dd S_\mathrm{YM}}{\dd A^a_i(t, \mx)}\right\vert_{\theta} + \eta^a_i(\theta, \lambda, \mx). 
        \end{split}
        \end{align}

    To relate \eq \eqref{eq:cle_ym_contour_hybrid} to the aforementioned Minkowski and Euclidean cases, one needs kernel freedom. The notion of kernels discussed in \se\ref{sec:param_contour} can be generalized to gauge theories. In particular, a kerneled CL equation for gauge theories is of the form
        \begin{align} \label{eq:ym_cle_kernel}
            \partial_\theta A^a_\mu(\theta, x) &= 
            i \int d^{d+1}x' \, \Gamma^{ab}_{\mu\nu}(x,x') \left. \frac{\delta S}{\delta A^b_\nu(x')} \right\vert_{\theta} + \int d^{d+1}x' \, \tilde \Gamma^{ab}_{\mu\nu}(x,x') \eta_b^\nu(\theta, x'), 
        \end{align}
        where the kernel $\Gamma^{ab}_{\mu\nu}(x,x')$ is factorizable
        \begin{align}
            \Gamma^{ab}_{\mu\nu}(x,x') = \int d^{d+1}x'' \, \tilde \Gamma^{ac}_{\mu\sigma}(x, x'') \tilde \Gamma^{bc}_{\nu\sigma}(x', x'').
        \end{align}
        Analogously to \se\ref{sec:param_contour}, kerneled CL equations are equivalent to the original CL equation for the gauge degrees of freedom. Hence, the kernel freedom allows us to effectively rescale all degrees of freedom of our model independently.

    As detailed in \app\ref{app:ym_contour_param}, the parameterized CL equation in \eq\eqref{eq:cle_ym_contour_hybrid} retains the stationary solutions of the previous formulation for Minkowski time contours in \eq\eqref{eq:ym_cle_mink} and Euclidean contours in \eq \eqref{eq:ym_cle_eucl} by virtue of the kernel freedom and by suitably relating the noise terms. 
    Similar to our previous toy model, kernel freedom also implies general reparameterization invariance of the stationary solution.

    \subsection{Discretizing the CL equation for Yang-Mills theory} 
    \label{sec:lattice}
        
    Following the same approach as in \se\ref{sec:toy_model}, the next step is to discretize the contour parameterized CL equation for gauge fields. We proceed by approximating the underlying space-time as a discrete lattice. For the spatial part we choose a regular cubic lattice with spacing $a_s$ for each time slice. Each spatial lattice site can thus be represented by a vector $\mx = a_s \sum^3_{i=1} n_i \hat e_i$ with unit vectors $\hat e_i$ parallel to the lattice axes and integer-valued coordinates $n_i \in \{0, 1, \dots, N_s-1\}$, where $N_s$ denotes the number of spatial sites in one direction. Since we consider general complex time contours $t(\lambda)$, we choose an inhomogeneous discretization for the time direction. The discrete time contour is defined by $t_k = t(\lambda_k)$ for $k \in \{0, 1, \dots N_t \}$ with a discrete contour parameter $\lambda_k$. Here, $N_t$ is the number of temporal lattice sites along the contour. The time and parameter spacings are given by $a_{t,k} = t_{k+1} - t_k$ and $a_{\lambda,k} = \lambda_{k+1} - \lambda_k$ for $k<N_t$. We use periodic boundary conditions for the degrees of freedom. 
    
    In contrast to our toy model, more care has to be taken regarding the degrees of freedom, namely the gauge fields $A_\mu^a(x)$. Lattice gauge theory provides a gauge symmetric discretization scheme, which uses group-valued gauge links as degrees of freedom. The set of gauge links $\{ U_{x,\mu} \}$ is given by Wilson lines along the lattice edges, 
    \begin{align} \label{eq:link_continuum}
        U_{x,\mu} &\approx \exp \left( i g a_\mu A_\mu\left(x+\frac{1}{2} \hat \mu\right) \right) \in \mathrm{SL}(N_c, \mathbb C)\,,
    \end{align}
    i.e.~connecting neighboring lattice sites $x$ and $x+a_\mu \hat e_\mu$ (no summation over $\mu$), and where links with a negative index $U_{x,-\mu}$ point into the opposite direction, i.e.~$U_{x,-\mu} = U_{x-\mu, \mu}^{-1}$. Thus, our goal is to recast the continuous CL evolution in \eq\eqref{eq:cle_ym_contour_hybrid} in terms of gauge links. Although our approach closely follows  \cite{Berges:2006xc}, there are a few subtleties regarding the discretized time contour, which motivate a careful re-derivation of the CL equations for gauge links. Here, we only summarize the results and refer to \app \ref{app:lattice} for a detailed derivation.

    Let us first introduce plaquettes $U_{x,\mu\nu}$ defined as $1 \times 1$ Wilson loops
    \begin{align}
        U_{x,\mu\nu} = U_{x,\mu} U_{x+\hat\mu, \nu} U_{x+\hat\nu, \mu}^{-1} U_{x,\nu}^{-1}\,,
    \end{align}
    which, in the limit of small lattice spacings, approximate the field strength tensor at the mid-point of the face spanned by directions $\hat\mu$ and $\hat\nu$
    \begin{align}
        U_{x,\mu\nu} \approx \exp \left( i g a_\mu a_\nu \,F_{\mu\nu}\left(x+\frac{1}{2} \hat\mu + \frac{1}{2} \hat \nu \right) \right).
    \end{align}
    This approximation allows us to discretize the Yang-Mills action in \eq\eqref{eq:ym_action} in terms of plaquettes, which yields the Wilson action
    \begin{align} \label{eq:wilson_action_compact}
        S_\mathrm{W}[U] = \frac{1}{2N_c} \sum_{x, \mu \neq \nu} \beta_{x,\mu\nu} \mathrm{Tr}\left[U_{x,\mu\nu} - 1\right],
    \end{align}
    with the coupling constants
    \begin{align}
        \beta_{x,0i} &=\beta_{x,i0}=-\beta_{k,0}=-\frac{2N_c}{g^2}\frac{a_s}{a_{t,k}}, \\
        \beta_{x,ij}&=\beta_{x,ji}=+\beta_{k,s}=+\frac{2N_c}{g^2}\frac{\bar a_{t,k}}{a_s}\,,
    \end{align}
    and the averaged time-step in the spatial plaquette term
    \begin{align}
        \label{eq:avg_timestep}
        \bar a_{t,k} = \frac{1}{2}\left(a_{t,k} + a_{t,k-1}\right) = \frac{1}{2}\left(t_{k+1} - t_{k-1}\right).
    \end{align}
    The latter guarantees time reversal symmetry and quadratic accuracy of the Wilson action even for general time contours. 

    The drift term entering the CL equation is the functional derivative of the Wilson action and reads (see \app \ref{app:lattice})
    \begin{align}
        \frac{\delta S_\mathrm{W}}{\delta \tilde A^a_{x,\mu}} = \frac{i}{2N_c} \sum_\nu \mathrm{Tr} \left[ t^a \left( \beta_{x,\mu\nu} \left(U_{x,\mu\nu} - U^{-1}_{x,\mu\nu} \right) + \beta_{x-\nu, \mu\nu} \left(U_{x,\mu-\nu} - U^{-1}_{x,\mu-\nu} \right) \right) \right],
    \end{align}
    where $t^a$ are the traceless Hermitian generators of the SU($N_c$) group and $\tilde A^a_{x,\mu} = g a_\mu A^a_{x,\mu}$. Taking all subtleties into account, we transform the CL equation in \eqref{eq:cle_ym_contour_hybrid} to an update equation for gauge links on complex time contours in \app \ref{app:lattice}, arriving at
    \begin{align} \label{eq:ym_contour_cle_unkerneled_temp}
        U_{x,t}( \theta +  \epsilon) &= \exp \left( i t^a \left[ i  \epsilon\, \frac{a_{\lambda,k}}{a_s}  \left.\frac{\delta S_\mathrm{W}}{\delta  \tilde A^a_{x,t}} \right\vert_{ \theta} + \sqrt{ \epsilon}\, \sqrt{\frac{a_{\lambda,k}}{a_s}}\, \eta^a_{x,\lambda}( \theta) \right] \right) U_{x,t}( \theta), \\ \label{eq:ym_contour_cle_unkerneled_spat}
        U_{x,i}( \theta +  \epsilon) &=  \exp \left( i t^a \left[ i  \epsilon\, \frac{a_s}{\bar a_{\lambda,k}}   \left. \frac{\delta S_\mathrm{W}}{\delta \tilde A^a_{x,i}} \right\vert_{ \theta} + \sqrt{ \epsilon} \, \sqrt{\frac{a_s}{\bar a_{\lambda,k}}}\, \eta^a_{x,i}( \theta) \right] \right) U_{x,i}( \theta)\,,
    \end{align}
    with the Langevin time step $\epsilon$ and the dimensionless noise correlator
    \begin{align} \label{eq:ym_cle_correlator}
        \langle \eta^a_{x,\mu}(\theta) \eta^b_{x',\nu}(\theta') \rangle &= 2 \delta_{\mu\nu} \delta^{ab} \delta_{\theta\theta'} \delta_{kk'} \delta_{\mx, \mx'}.
    \end{align}
    
    \subsection{Comparison to earlier approaches} 
    \label{sec:early_ym_cle}

    Our re-derivation of the CL equations for gauge links shows that for a general time contour, additional prefactors in front of the drift and noise terms appear. It is then instructive to compare our update step to the one used in \cite{Berges:2006xc}, which in our notation reads
    \begin{align} 
    \label{eq:berges_cl}
    U_{x, \mu}(\theta+\epsilon) &= \exp \left( i t^a \left[ 
     i \epsilon  \left.\frac{\dd S_\mathrm{W}}{\dd \tilde A^a_\mu}\right\vert_{\theta} + \sqrt{ \epsilon}\, \eta^a_{x,\mu}(\theta)
     \right] \right) U_{x, \mu}(\theta),
    \end{align}
    where the noise term satisfies the same correlator as in \eq\eqref{eq:ym_cle_correlator}. Comparing this update step to \eqs\eqref{eq:ym_contour_cle_unkerneled_temp} and \eqref{eq:ym_contour_cle_unkerneled_spat}, we see that the main differences are contour and lattice spacing dependent prefactors. Similarly to our discussion at the end of \se\ref{sec:ym_contour_param}, these can be absorbed into a CL kernel. 
    Therefore, both schemes are equivalent due to the kernel freedom provided that a convergent solution is reached for $\theta \rightarrow \infty$. Additionally, our update scheme reduces to \eq\eqref{eq:berges_cl} if we choose a homogeneously discretized contour with $a_{\lambda,k} = \bar a_{\lambda,k} = a_s$.

    In \re\cite{Berges:2006xc} it was reported that the CL update procedure in \eq\eqref{eq:berges_cl} is prone to an instability towards wrong results, which we will refer to as \emph{wrong convergence} in the following.
    The severity of such an instability is tightly correlated to the value of the tilt angle $\alpha$ of the complex time contour in \fig\ref{fig:contour}. The smaller the angle, the earlier one finds wrong convergence, as we will discuss in \se\ref{sec:results}. This makes it impossible to obtain correct results for sufficiently small tilt angles with the update step of \eq\eqref{eq:berges_cl}.
    
    To mitigate this unstable behaviour, we investigate known stabilization methods that were designed to improve convergence and stability in CL simulations of QCD at finite chemical potential. 
    In particular, we conduct thorough tests of adaptive stepsize (AS) \cite{Aarts:2009dg}, gauge cooling (GC) \cite{Seiler:2012wz} and dynamical stabilization (DS) \cite{Attanasio:2018rtq}. Their impact on the CL dynamics for real-time Yang-Mills theory will be shown in \se\ref{sec:results_stability} while details of these stabilization techniques and of how we use them are revisited in \app\ref{sec:stabilization}. 
    Consistent with \cite{Aarts:2017hqp}, our conclusion is that while they improve the simulations, they are insufficient to obtain stable and correct results for small enough tilt angles. 
    Therefore, another approach is required to obtain correct expectation values. In the next subsection we motivate the introduction of a novel anisotropic kernel which shows great potential by systematically improving on the stability of our simulations as shown in \se\ref{sec:results}. 
    
    \subsection{Regularization of the path integral and a new anisotropic kernel}
    \label{sec:new_method}

    We will now introduce another approach to stabilize real-time CL simulations which is well motivated by the regularization of the path integral for lattice gauge theories. It was pointed out that the path integral for the Wilson action on a Minkowski time contour has an ill-defined continuum limit and thereby does not yield a unitary theory \cite{Hoshina:2020gdy, Kanwar:2021tkd, Matsumoto:2022ccq}. It was proposed in \re\cite{Kanwar:2021tkd} to resolve this issue through path integral contour deformation. The ill-defined continuum limit for the Wilson action was traced back to the character expansion of compact variables in terms of modified Bessel functions in \re\cite{Matsumoto:2022ccq}.
    
    A possible way to regularize the path integral on the Minkowski contour is to multiply the lattice-spacing dependent coupling constant by a phase factor \cite{Matsumoto:2022ccq}
    \begin{align}
    	\beta_0  &\rightarrow e^{+i\alpha} \beta_0, \\
    	\beta_s  &\rightarrow e^{-i\alpha} \beta_s,
    \end{align}
    where $\alpha > 0$. In the context of real-time simulations, this regularization can be understood as an infinitesimal Wick rotation of the positive branch $\mathscr{C}^+$ of the Schwinger-Keldysh time contour. For the negative branch $\mathscr{C}^-$ one has to invert the phase factor $e^{+i\alpha} \rightarrow e^{-i\alpha}$.
    The resulting tilted contour is visualized as an orange line in \fig\ref{fig:disc_vs_cont_contour}. This prescription translates to the following replacements for the temporal lattice spacing along the time contour:
    \begin{alignat}{3}
    	a_{t,k} &\rightarrow e^{-i\alpha} a_{t,k} &\quad \text{for} \quad &t_k \in\mathscr{C}^+, \\
    	a_{t,k} &\rightarrow e^{+i\alpha} a_{t,k} &\quad \text{for} \quad &t_k \in\mathscr{C}^-.
    \end{alignat}
    The regularization parameter $\alpha$ then corresponds to the tilt angle of the contour. In \cite{Matsumoto:2022ccq} it was further emphasized, that the order of limits of the temporal continuum limit $a_{t,k} \to 0$ and the subsequent limit to small tilt angles $\alpha \to 0$ has to be respected to ensure the convergence of the path integral in \eq\eqref{eq:pi_gauge}.
    
    The introduced CL formalism for complex time paths via contour parametrization can be utilized to account for this regularization by simply using a tilted Schwinger-Keldysh contour with tilt angle $\alpha$. Hence, we implicitly introduce the phase factors in the update steps in \eqs\eqref{eq:ym_contour_cle_unkerneled_temp} and \eqref{eq:ym_contour_cle_unkerneled_spat} which are not explicitly dependent on $\alpha$. An interesting aspect of our update scheme is that issues may arise when we undo the discretization of the time contour and take the continuous time limit. More specifically, taking the limit $a_{t,k} \sim a_{\lambda,k} \rightarrow 0$ for fixed spatial lattice spacing $a_s$ and fixed Langevin step $ \epsilon$ is potentially problematic due to the blow-up of the factors $a_s / \bar a_{\lambda,k}$ in the spatial update step.
    
    Fortunately, we can exploit the kernel freedom of CL to circumvent this problem. The main idea is to rescale the Langevin time step for the spatial update in \eq\eqref{eq:ym_contour_cle_unkerneled_spat} via
    \begin{align}
    	\epsilon \rightarrow \epsilon_{k,s} \equiv \epsilon \frac{\bar a_{\lambda, k}}{a_s},
    \end{align}
    which for $a_{\lambda,k} < a_s$ effectively slows the Langevin evolution of the spatial links. Similarly, we may perform an analogous rescaling for the temporal update in \eq\eqref{eq:ym_contour_cle_unkerneled_temp} via
    \begin{align}
    	\epsilon \rightarrow \epsilon_{k,0} \equiv \epsilon \frac{a_{\lambda, k}}{a_s}.
    \end{align}
    This contour-dependent rescaling of $\epsilon$ can be understood as a field-independent kernel transformation applied to our contour-parameterized CL update scheme.%
    \footnote{We thank Daniel Alvestad for pointing out this correspondence.}
    The new kerneled update steps are then given by
    \begin{align} \label{eq:new_cle_temp}
    	U_{x,t}( \theta +  \epsilon) &= \exp \left( i t^a \left[ i  \epsilon \left(\frac{a_{\lambda, k}}{a_s}\right)^2  \left.\frac{\delta S_\mathrm{W}}{\delta  \tilde A^a_{x,t}} \right\vert_{ \theta} + \sqrt{ \epsilon}\, \frac{a_{\lambda, k}}{a_s}\, \eta^a_{x,\lambda}( \theta) \right] \right) U_{x,t}( \theta), \\ \label{eq:new_cle_spat}
    	U_{x,i}( \theta +  \epsilon) &=  \exp \left( i t^a \left[ i  \epsilon  \left. \frac{\delta S_\mathrm{W}}{\delta \tilde A^a_{x,i}} \right\vert_{ \theta} + \sqrt{ \epsilon}\, \eta^a_{x,i}( \theta) \right] \right) U_{x,i}( \theta),
    \end{align}
    which is well-behaved in the limit of $a_{\lambda, k} \rightarrow 0$. Comparing this kerneled update scheme to the more traditional scheme of \eq\eqref{eq:berges_cl}, we see that our modification amounts to an \emph{anisotropic kernel}, which treats temporal and spatial links differently. More specifically, the Langevin update for temporal links is slowed down by a factor of $(a_{\lambda,k} / a_s)^2$ compared to the update of spatial links. Similar to the toy model in \se\ref{sec:toy_model}, we choose $a_{\lambda, k} = \left| a_{t,k} \right|$ for practical simulations, although any other parameterization would also be admissible. 

    In \se\ref{sec:systematics} we show that increasing the number of temporal lattice sites $N_t$, and thereby decreasing the temporal lattice spacing $a_{t,k}$, leads to improved stability of our simulations. We want to highlight that this systematic behaviour is precisely what is needed to obtain correct results for the continuum limit for the following reasons. We have already remarked that the correct order of limits is followed by calculating $a_{t,k} \to 0$ of the path integral first for finite $\alpha$ and subsequently taking the limit to vanishing tilts $\alpha \to 0$. In practice we can utilize the continuum limit via the anisotropic kernel to reach better stability of the simulation, which at the same time allows us to simulate for decreasing tilt angles.
    Our approach is therefore well-suited to sample the path integrals of {SU($N_c$)} gauge theories on the Schwinger-Keldysh contour. 

\section{Stabilizing real-time Yang-Mills simulations} 
    \label{sec:results}
    
    We test the impact of our kernel on the stability of the CL method by simulating SU(2) Yang-Mills theory on a 3+1 dimensional lattice. To reach improved stability we utilize modern stabilization techniques summarized in \app\ref{sec:stabilization} in conjunction with our novel anisotropic kernel \new{in the CL update step} \eqref{eq:new_cle_temp} and \eqref{eq:new_cle_spat}. Compared to the traditional update step of \eq\eqref{eq:berges_cl}, our kernel introduces an explicit lattice spacing dependence of our CL step. This allows us to systematically improve the stability of the simulations by increasing the number of temporal lattice sites $N_t$.

    The use of stabilization methods combined with our anisotropic kernel alleviates problems of wrong convergence to a large degree, but does not fully resolve them. In particular, we are not able to achieve correct convergence for arbitrarily large Langevin times $\theta \rightarrow \infty$. Similar to what has been demonstrated in \cite{Berges:2006xc}, the general behaviour of the CL evolution is as follows: the stochastic process first undergoes a transition away from the initial state towards a region of correct convergence. Depending on the physical and numerical parameters of the system, the evolution resides in this region for a certain time, which we term the \emph{metastable region}.
    However, as the evolution continues, we observe that the process moves towards another stable solution that exhibits wrong convergence, i.e.~the expectation values eventually converge to wrong results. This particular instability, which drives the process away from the region of correct convergence, can be strongly reduced by our new update equations. We find that the anisotropic kernel allows us to systematically extend the metastable region by increasing the resolution along the complex time contour. Hence, we are able to extract correct expectation values by taking samples only from the metastable region, which can in principle be made arbitrarily large at the cost of an increased lattice size. \new{Additional statistics can be gained by sampling over independent simulations. However, we will not follow this strategy in this paper since here our goal is to assess the CL method with our new kernel.}

    \subsection{Numerical setup}
     To compare our results with \re\cite{Berges:2006xc}, we consider an isosceles contour as depicted in \fig\ref{fig:isosceles_contour}. The forward and backward parts $\mathscr{C}^+$, $\mathscr{C}^-$ of the isosceles contour connect $t=0$ with $t=\max[\Re(t)]-i\beta/2$  and  $t=\max[\Re(t)]-i\beta/2$ with $t=-i\beta$ respectively, where $\beta$ is the inverse temperature\footnote{The inverse temperature $\beta$ should not be confused with the couplings $\beta_{x,\mu\nu}$ in the Wilson action.}.
     The maximal real-time extent $\max[\Re(t)]$ of the isosceles contour can be related to the tilt angle $\alpha$ via
    \begin{align} \label{eq:tilt_angle}
        \tan\left(\alpha\right) = \frac{\beta/2}{\max[\Re(t)]}.
    \end{align}
    The Euclidean contour is realized in the limit $\alpha \rightarrow \pi/2$ where $\tan\left(\alpha\right) \rightarrow \infty$.  Large but finite values of $\tan\left(\alpha\right)$ lead to small real-time extents, whereas for $\tan \alpha \leq 0.5$, the real-time extent exceeds the inverse temperature $\max[\Re(t)] \geq \beta$. Although we only present results for the isosceles contour, we note that with our new update scheme, we observed similar improvements of stability for discretized Schwinger-Keldysh contours depicted as the orange line in \fig\ref{fig:disc_vs_cont_contour}. 
    
    The calculation of expectation values using the CL method is achieved by sampling the observable from the stochastic process of the link configuration. We initialize our simulations with homogeneous configurations of identity matrices%
    \footnote{The choice of initial conditions for the CL evolution is not unique. In our numerical experiments we have also tested initial conditions with random unitary gauge links and found qualitatively similar behaviour.} 
    and evolve them sufficiently until the system equilibrates. Once a metastable equilibrium is reached, observables fluctuate around constant values. 
    \new{In practice, we use the time-translation invariance of thermal equilibrium to estimate the boundaries of the metastable region. As we will show below, we compare a time-independent observable $\mathscr{O}$ on the complex time path to the results from a (stable) Euclidean simulation where the value can be calculated to high accuracy. For instance, we set the beginning and the end of the metastable region at the points when the observable roughly coincides with the Euclidean simulation and when a plateau has formed around which the observable fluctuates. We have checked that for a sufficiently long metastable region, our exact choice for its starting and ending Langevin times becomes less important.}
    
    In general, observables at different Langevin times are correlated. To obtain unbiased expectation values, we sample data points that are sufficiently far apart with respect to the autocorrelation (Langevin) time. We approximate the autocorrelation function $R_{\mathscr{O}}$ associated with an observable $\mathscr{O}$ with
    \begin{align}
        R_{\mathscr{O}}(\tau) &= 
        \frac{
            \langle 
            \left( \mathscr{O}_{\theta}- \langle \mathscr{O}_{\theta} \rangle \right)  \left( \mathscr{O}_{\theta + \tau}- \langle \mathscr{O}_{\theta + \tau} \rangle \right) \rangle
        }
        {\sigma_{\theta} \sigma_{\theta + \tau}} \approx \exp\left(-\tau  / T_{\mathscr{O}}\right),
    \end{align}
    and find the autocorrelation time $T_{\mathscr{O}}$ by a parameter fit with an exponential decay. The standard deviation of the observable at Langevin time $\theta$ is denoted by $\sigma_{\theta}$. In \new{most} figures of this section where we show the evolution of an observable over $\theta$, we rescale the Langevin time by the autocorrelation time. This allows a fair comparison of different CL simulations. Additionally, we use simple moving averages to reduce the overall noise of our data, which allows us to better see the systematic behaviour in our plots.

    We consider systems with inverse temperature $\beta = 4$ in lattice units, spatial lattice spacing $a_s = 1$ and coupling constant $g=1$ on a 3+1 dimensional lattice with $N_s=4$ and $N_t=16$ if not stated otherwise. \new{In our current setup the temporal lattice spacing $a_t$ is chosen such as to match a given $N_t$, tilt angle and $\beta$ in lattice units. For the stated standard values this implies a bare spacing of $a_t = a_s / 4$ for a Euclidean contour.}
    The stabilization methods discussed in \app\ref{sec:stabilization} have additional numerical parameters that must be chosen appropriately. We set these parameters by hand to optimize the stability of the simulations while keeping possible biases small. We use a bound parameter $B=(2\pi)^3$ for the adaptive step size (AS), which removes runaway instabilities without distorting the results. For gauge cooling (GC), the number of gauge cooling steps is chosen between one and five with a force parameter within $\alpha_{GC} \in [10^{-4}, \, 5 \cdot 10^{-3}]$. For dynamical stabilization (DS) we observed that the parameter $\alpha_{DS} \in [1, 10^{3}]$ leads to a stabilizing effect for sufficiently large tilt angles $\alpha$, but introduces a small bias in the obtained results. We observed that we lose the stabilizing effect for smaller $\alpha_{DS}$ and introduce a significant bias for larger $\alpha_{DS}$. \new{Note that DS always introduces a bias due to its modifications to the CL equation, as will become evident from \mbox{Fig.~\ref{fig:avg_spat_plaq_a} and Table \ref{tab:avg_spat_plaq}}. We will therefore refrain from combining this stabilization method with our kernel and only use it for comparisons.}

    \subsection{Effectiveness of the new CL update scheme}
    \label{sec:results_stability}
    
        We assess the effect of the introduction of the anisotropic kernel in \eqs\eqref{eq:new_cle_temp} and \eqref{eq:new_cle_spat} by calculating expectation values of time-independent, gauge invariant observables such as the average spatial plaquette%
        \footnote{\new{We have checked that we obtain similar starting and end points of the metastable region for other observables like the expectation values of $2 \times 2$ and $3 \times 3$ Wilson loops.}}
        \begin{align} \label{eq:avg_spat_plaquette}
            \mathscr{O}[U]=\frac{1}{N_t N_s^3}\sum_x \frac{1}{3} \sum_{i < j} \frac{1}{N_c} \mathrm{Re} \mathrm{Tr} \, U_{x,ij}.
        \end{align}
        We compare our simulation results with those obtained from the traditional update scheme \eq\eqref{eq:berges_cl} originally used in \cite{Berges:2006xc}.
        Focusing on time-independent observables has the remarkable benefit that they can be computed from real Langevin simulations on Euclidean contours where no sign problem occurs and the gauge links remain unitary. As a result, the Euclidean simulations do not exhibit any instabilities towards wrong convergence and reliably produce reference data that allows us to check the validity of the CL results. 
        
\begin{figure}[t]
            \centering
            \begin{subfigure}[t]{.49\textwidth}
                \centering\includegraphics[width=1\linewidth]{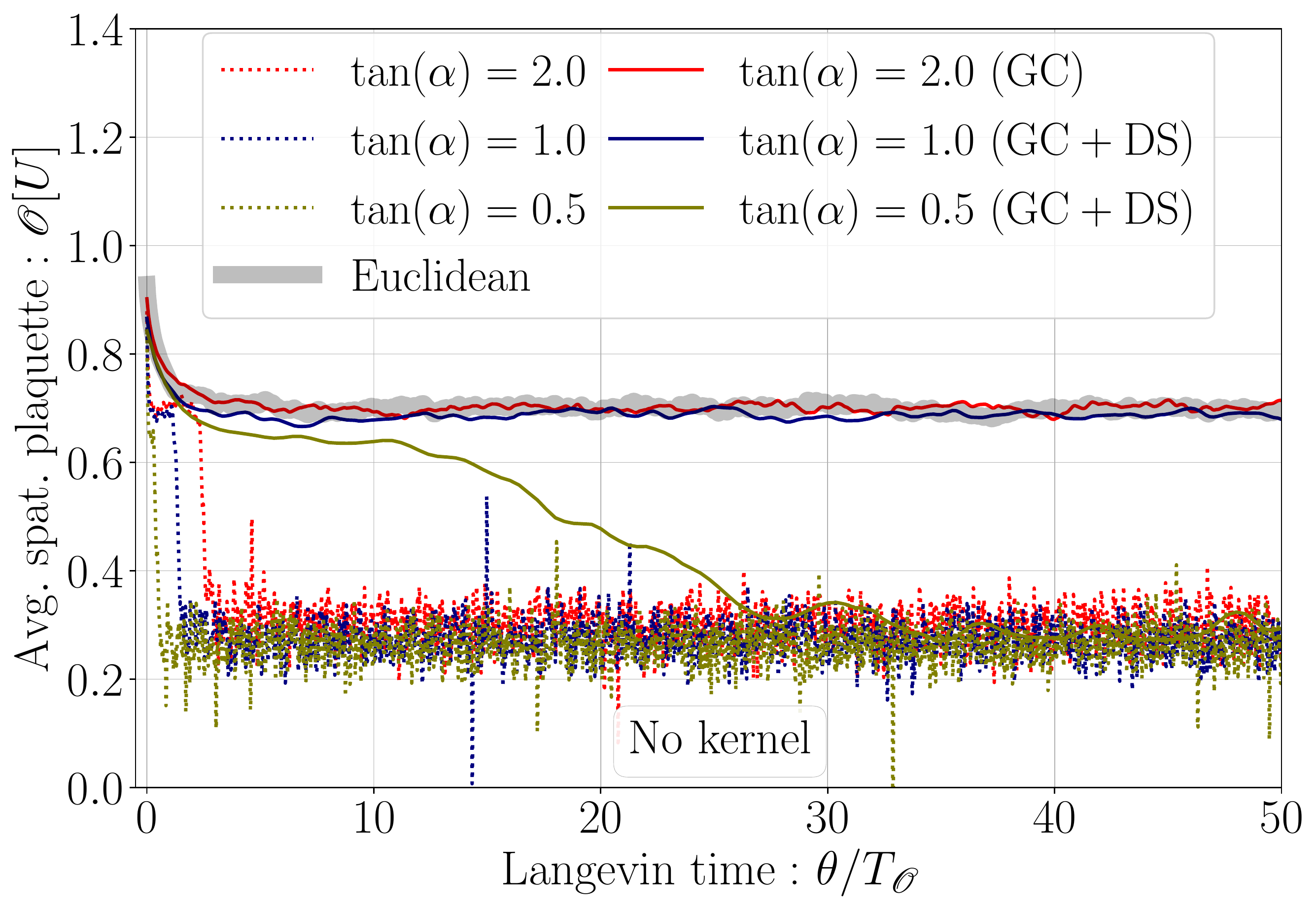}
                \caption{\label{fig:avg_spat_plaq_a}}
            \end{subfigure}
            \begin{subfigure}[t]{.49\textwidth}
                \centering\includegraphics[width=1.\linewidth]{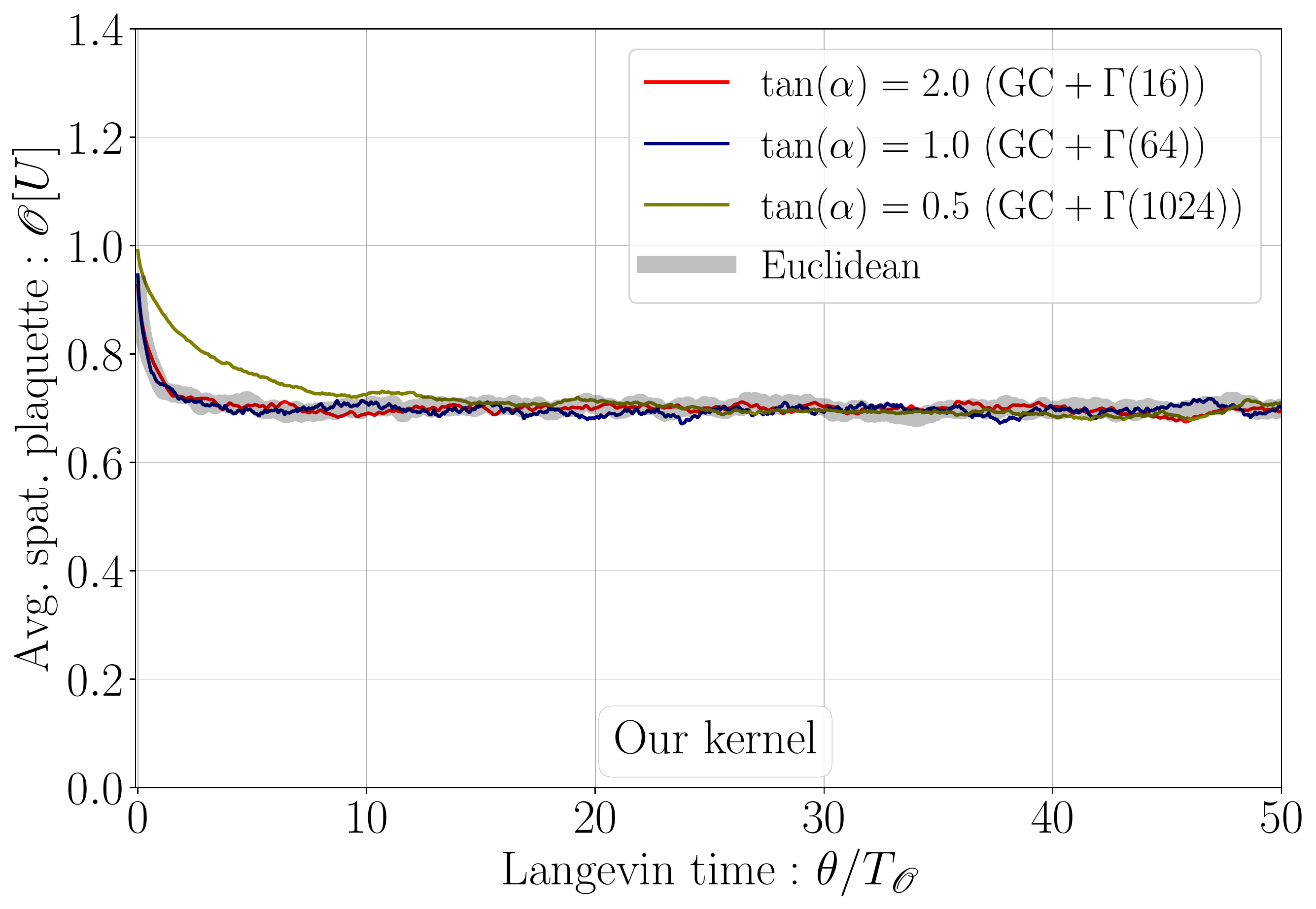}
                \caption{\label{fig:avg_spat_plaq_b}}
            \end{subfigure}
            \caption{(a) Average spatial plaquette $\mathscr{O}$ as a function of Langevin time $\theta$ from Euclidean simulations and CL simulations using various tilt angles $\alpha$ for a complex isosceles contour.
            Dotted lines show simulations without stabilization and solid lines show results for various stabilization techniques. (b) Average spatial plaquette $\mathscr{O}(\theta)$ using our anisotropic kernel denoted by $\Gamma(N_t)$. {\em (Both panels)} For simulation which converge to the correct result, the Langevin time is shown in units of the autocorrelation time of the sampled data. 
            The interval $\theta / T_\mathscr{O} \in [0, 50]$ is shown to resolve instabilities towards wrong convergence, however simulations are evolved up to $\theta / T_\mathscr{O} = 100$. 
            \label{fig:avg_spat_plaq}}
        \end{figure}

\begin{figure}[t]
            \centering
            \centering\includegraphics[width=0.5\textwidth]{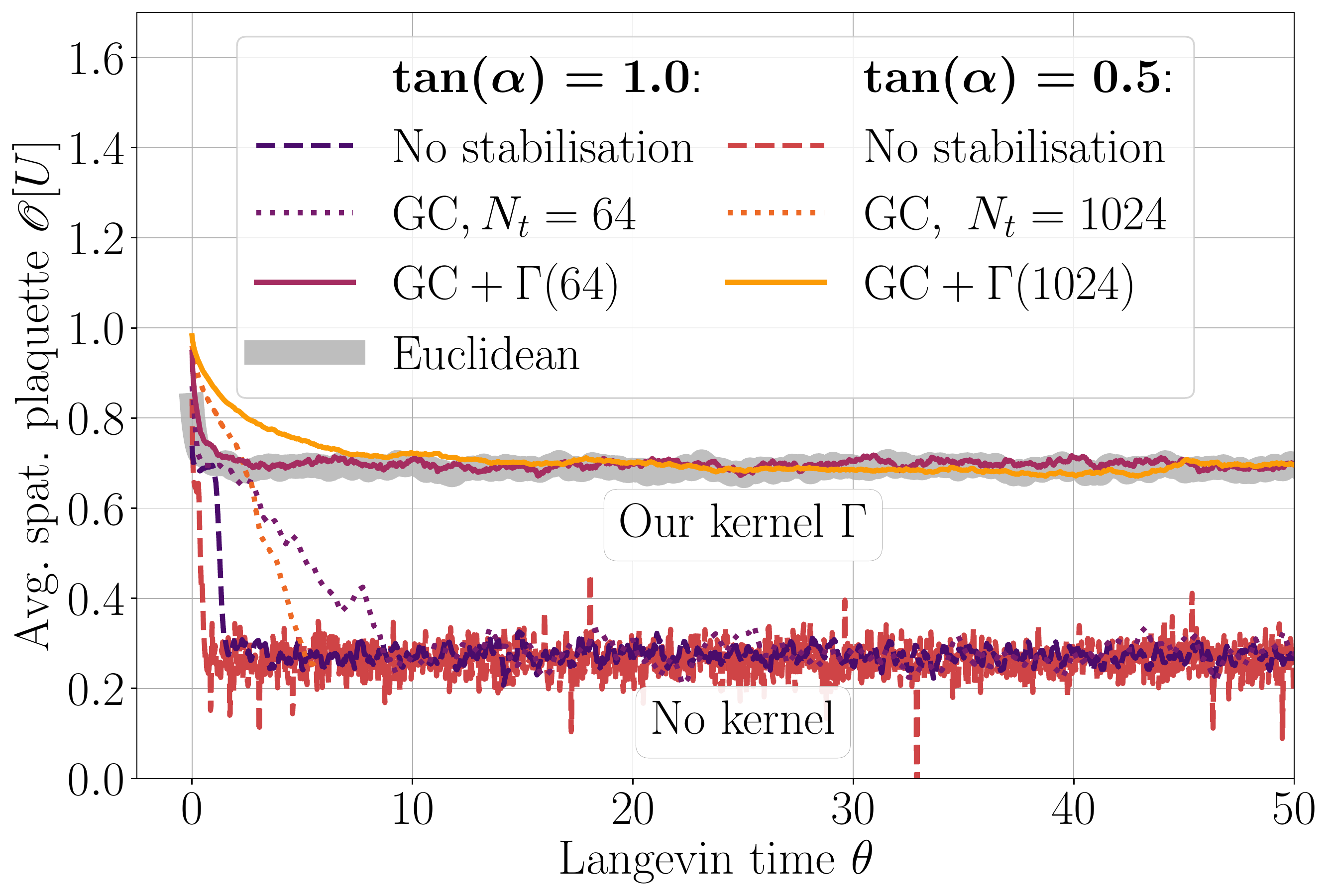}
            \caption{
            \new{Average spatial plaquette $\mathcal{O}$ for contour tilt angles $\alpha$ with an adaptive step size and (i) no further stabilization, (ii) with GC and a higher temporal resolution, and (iii) with GC and our kernel $\Gamma(N_t)$ for the same temporal resolution. 
            In the figure, convergence to the Euclidean result (gray) is only achieved with our kernel. Merely increasing the temporal resolution does not lead to correct convergence.
            }
            \label{fig:gc_vs_gcandgamma}}
        \end{figure}

        The numerical results of our simulations are summarized in \figs \ref{fig:avg_spat_plaq}, \ref{fig:gc_vs_gcandgamma} and \tab\ref{tab:avg_spat_plaq}. 
        \new{In the table,} the expectation values are computed in the metastable region after a short transition time. We take samples in half-steps of the autocorrelation time $T_\mathscr{O}$ to calculate the mean values and the statistical errors. In the case of unstable simulations, where the autocorrelation time becomes very small due to large fluctuations, we simply average over the whole ensemble of samples.
        
        \begin{table}[tp]
            \centering
            \begin{tabular}{M{0.2\textwidth} | M{0.3\textwidth} M{0.1\textwidth} M{0.2\textwidth}} \toprule
            {$\tan(\alpha)$} & {Stabilization techniques} & {$N_t$} & {$\langle\mathscr{O}\rangle$} \\ \midrule
            {(Euclidean)} & {None} & {16} & {$0.6992(3)$} \\ 
            {(Euclidean)} & {None} & {64} & {$0.6979(8)$} \\
            {(Euclidean)} & {None} & {1024} & {$0.6974(6)$} \\
            {(Euclidean)} & {None} & {8192} & {$0.6971(2)$} \\ \midrule
            {2.0} & {AS, GC} & {16} & {$0.6981(2)$} \\ 
            {1.0} & {AS, GC, DS} & {16} & {$0.6858(1)$} \\
            {0.5} & {AS, GC, DS} & {16} & {$0.2801(3)$} \\ \midrule
            {2.0} & {AS, GC, $\Gamma$} & {16} & {$0.6987(3)$} \\ 
            {1.0} & {AS, GC, $\Gamma$} & {64} & {$0.6977(5)$} \\ 
            {0.5} & {AS, GC, $\Gamma$} & {1024} & {$0.6973(8)$} \\
            {0.4} & {AS, GC, $\Gamma$} & {8192} & {$0.6968(6)$} \\ \bottomrule
            \end{tabular}
            \caption{Numerical results for expectation values $\langle\mathscr{O}\rangle$ of spatial plaquettes from Euclidean simulations and CL simulations on the complex isosceles contour of \fig\ref{fig:isosceles_contour}. 
            The first column shows the tilt angle $\alpha$. The second and third columns list the utilized stabilization techniques and the number of temporal lattice sites $N_t$. The fourth column contains the obtained numerical results including statistical uncertainty. In all cases GC is required to obtain stable results. DS can stabilize smaller tilt angles, but introduces a bias. \new{For all tested tilt angles, our anisotropic kernel (denoted by $\Gamma$) leads to unbiased stable results that agree with the Euclidean values for the same $N_t$ resolution to remarkable accuracy.} The expectation values are calculated by measuring $\mathscr{O}$ every half auto-correlation length $T_\mathscr{O}/2$ from the metastable region of one CL trajectory.
            \label{tab:avg_spat_plaq}} 
        \end{table}
        
        The CL simulations are carried out for different tilt angles $\alpha$ of the isosceles contour. In \re\cite{Berges:2006xc} it was shown that the severity of the instabilities towards wrong convergence regions depends on the tilt angle, as smaller tilts lead to an earlier transition towards the wrong distribution.
        The dotted lines in \fig\ref{fig:avg_spat_plaq_a} reproduce the values presented in \re\cite{Berges:2006xc}. We observe that the Langevin process initially approaches the correct value for the observable \new{for some angles} but then transitions and converges to another wrong expectation value.

        The transition towards wrong convergence can be mitigated in part by using stabilization methods such as gauge cooling (GC) and dynamical stabilization (DS). For a large tilt angle $\tan(\alpha)=2.0$ we see a promising stabilization effect when the configuration is iteratively gauge transformed via the GC procedure. However, for smaller tilts we found that GC is not sufficient and can even lead to numerical instabilities associated with the blow up of the gauge gradient. Confirming previous studies \cite{Aarts:2017hqp}, a combination of GC and DS can stabilize tilt angles down to $\tan(\alpha)=1.0$ but introduces a bias due to the DS penalty term, which leads to a slight offset of the solid blue line in \fig\ref{fig:avg_spat_plaq_a} and \new{of the corresponding} expectation value given in \tab\ref{tab:avg_spat_plaq} \new{when compared to the Euclidean value}. We further observe that DS breaks down for tilt angles as small as $\tan(\alpha)=0.5$ due to the rapid growth of the unitarity norm (as we will see in the next subsection), 
        which leads to a dominating penalty term. In this case, we found no admissible choice of the DS parameter $\alpha_{\mathrm{DS}}$ \new{such that DS stabilizes the simulation without significantly distorting the expectation value.}

        The results of our simulations with the anisotropic kernel are shown in \fig\ref{fig:avg_spat_plaq_b} and at the bottom of \tab\ref{tab:avg_spat_plaq}. In order to stabilize the simulations for small tilt angles, we increase the number of temporal lattice sites $N_t$.
        In addition, we perform gauge cooling after each update step, but do not use dynamical stabilization. As can be seen from the expectation values in \tab\ref{tab:avg_spat_plaq}, our anisotropic kernel yields results in \new{very} good agreement with the Euclidean case \new{with the same temporal resolution} for all tested tilt angles and, in contrast to the GC+DS simulations, does not induce a bias.

        It is interesting to point out that merely increasing the number of lattice sites $N_t$ without using our kernel does not lead to improvements regarding stability. 
        \new{This is shown in \mbox{Fig.~\ref{fig:gc_vs_gcandgamma}} for two tilt angles.}
        In both cases, we observe no improvement of stability by choosing finer discretizations of the time contour.
        \new{In contrast,} an extended metastable region
        is only achieved in the figure using the anisotropic kernel of \eqs\eqref{eq:new_cle_temp} and \eqref{eq:new_cle_spat} 
        \new{in combination with a higher temporal resolution.}

    \subsection{Dyson-Schwinger equations and unitarity norm}
        \label{sec:monitoring}
        
        In the previous section we have shown that we can successfully reproduce Euclidean data by stabilizing CL using the anisotropic kernel. In order to gain further trust in the simulation results, we check if the Dyson-Schwinger equations hold for the plaquette variables. For spatial plaquettes, this relation is given by (see \re\cite{Berges:2006xc})
        \begin{align} \label{eq:dse}
            \begin{split}
                \frac{2(N_c^2-1)}{N_c} \sum_{i<j} \left\langle \mathrm{ReTr}(U_{x, ij}) \right\rangle = 
                \frac{i}{2N_c} \sum_{i<j} \sum_{\vert \rho \vert \neq i} 
                \left\langle \mathrm{ReTr} 
                \left[ \beta_{x, i \rho} 
                ( U_{x,i\rho} - U_{x,i\rho}^{-1} ) U_{x,ij}
                \right] 
                \right\rangle.
            \end{split}
        \end{align}
        \begin{figure}[!t]
            \centering
            \centering\includegraphics[width=1\linewidth]{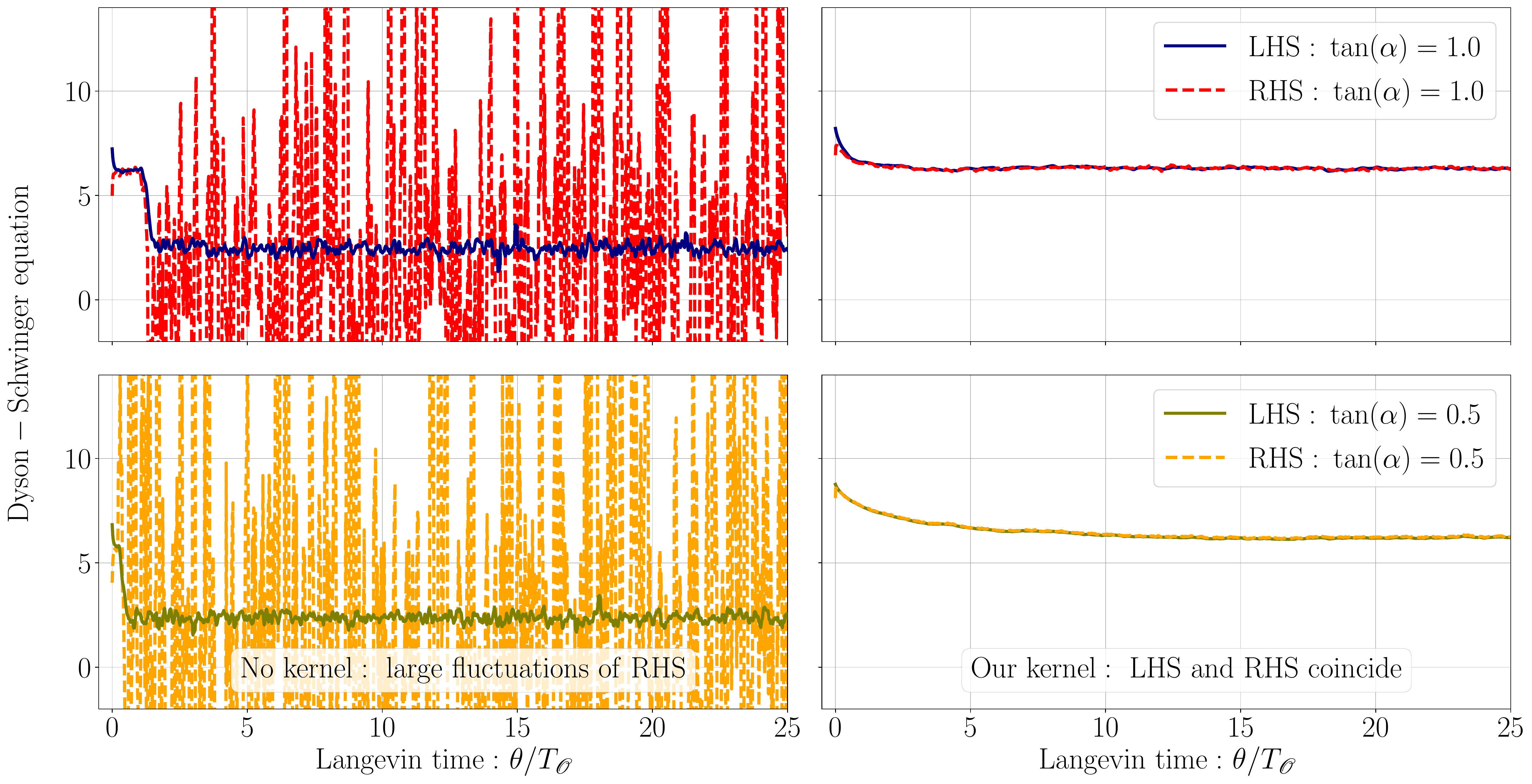}
            \caption{
            Results for the left- and right-hand side of the Dyson-Schwinger equations of the average spatial plaquette as function of the Langevin time in units of the autocorrelation time for isosceles contours with tilt angles $\tan(\alpha)=0.5$ and $\tan(\alpha)=1.0$. The same discretization as in Fig.~\ref{fig:avg_spat_plaq} was used in the simulations.
            (\emph{Left}) Values obtained by simulations with only AS.
            (\emph{Right}) Simulation results utilizing AS, GC and our anisotropic kernel.
            \label{fig:dse}}
        \end{figure}

        \begin{figure}[!t]
            \centering
            \centering\includegraphics[width=1\linewidth]{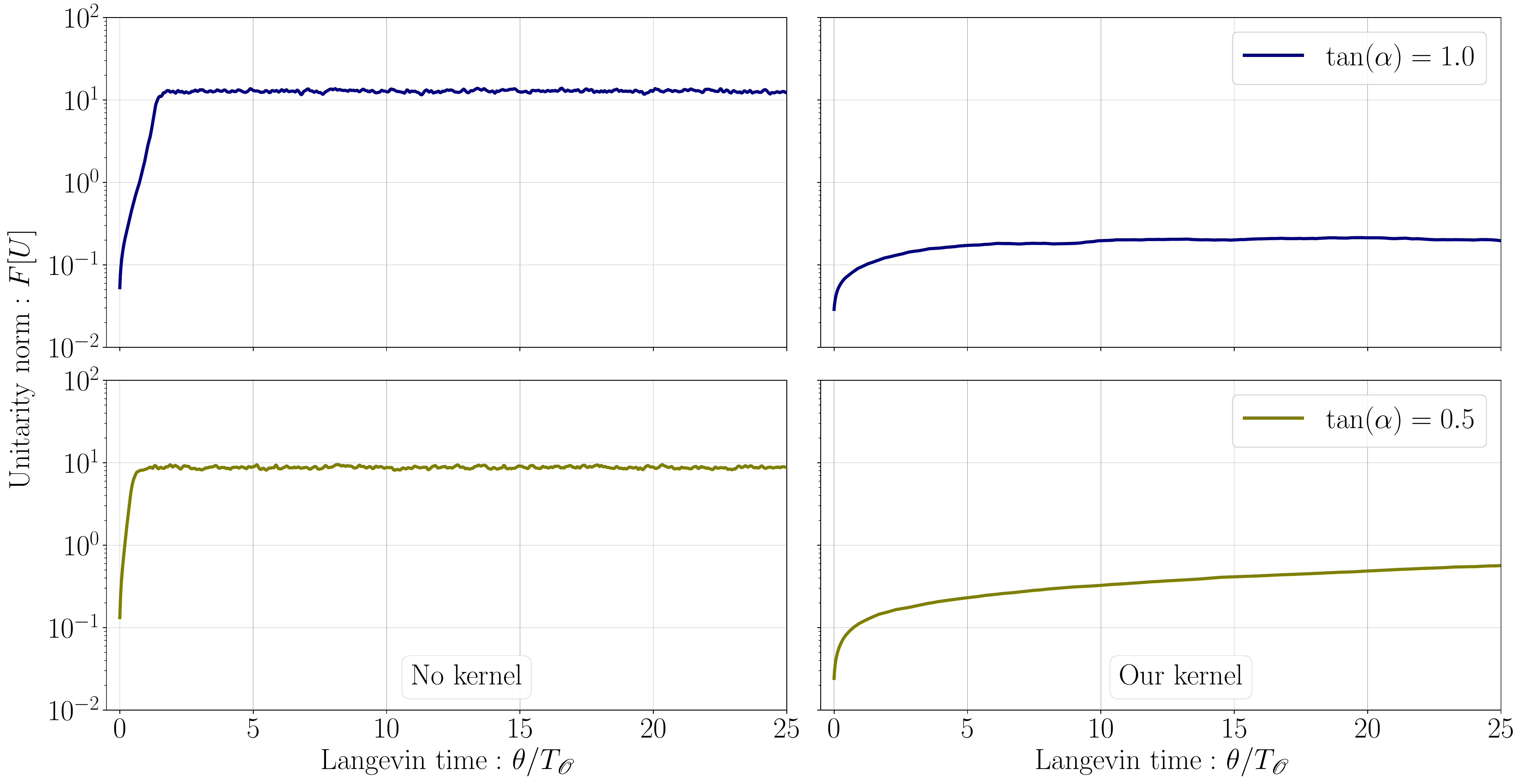}
            \caption{Unitarity norm $F[U]$ as a function of the Langevin time in units of the autocorrelation time for isosceles contours with tilt angles $\tan(\alpha)=0.5$ and $\tan(\alpha)=1.0$. The same discretization as in Fig.~\ref{fig:avg_spat_plaq} was used in the simulations.
            (\emph{Left}) Results obtained by simulations with only AS.
            (\emph{Right}) Simulation results utilizing AS, GC and our anisotropic kernel.\label{fig:unorm}}
        \end{figure}
        
        In \fig\ref{fig:dse} we show the numerical results of the left-hand side (LHS) and right-hand side (RHS) of \eq\eqref{eq:dse} as a function of Langevin time $\theta$.%
        \footnote{In order to obtain a simple measure to what degree the Dyson-Schwinger equations are satisfied, we average over all lattice sites and perform the expectation value for small ranges of Langevin time ($\Delta \theta \ll T_\mathscr{O}$).}
        We show simulations without stabilization in the left panels and results with our kernel in the right panels (as in \fig\ref{fig:avg_spat_plaq_b}).
        It has been remarked in \re\cite{Berges:2006xc} that the Dyson-Schwinger equations approximately hold even for the wrong convergence region. However, we find a clear difference between the regions of correct and wrong convergence regarding the RHS. 
        We observe that for regions of wrong convergence, as shown in the left panels of \fig\ref{fig:dse}, the RHS of the equation is governed by fluctuations around the same values as obtained for the LHS but with a much larger variance. 
        In contrast, stabilized simulations using GC and our kernel (right panels of \fig\ref{fig:dse}) indicate that both LHS and RHS coincide with little noise as long as we remain in the metastable region with correct convergence. 
        Remarkably, as can be seen from unstablized simulations with $\tan\left(\alpha\right) = 1.0$ in the upper left panel, the stochastic process initially transitions through a short lived metastable region with small fluctuations. 
        Since all of our simulations show the same pattern, we conclude that the fluctuations of the RHS of the Dyson-Schwinger equations can be taken as an indicator of a metastable region. 
        It has been demonstrated in simple models \cite{Scherzer:2018hid} that there is a connection between wrong convergence and the appearance of boundary terms.%
        \footnote{See also the discussion in \re\cite{Pehlevan:2007eq} about the relation between stationary solutions of CL equations and complexified solutions to the Dyson-Schwinger equations.} 
        Due to our results we therefore expect no boundary terms to emerge in the Dyson-Schwinger equations in the metastable region and plan to investigate this more carefully in the future.
        
        As an additional observable for validating our simulations, we compute the unitarity norm
        \begin{align} \label{eq:unorm_main}
            F[U] = \sum_{x,\mu} \mathrm{Tr}\left[(U_{x, \mu} U_{x, \mu}^\dagger - 1)^2\right] \geq 0.
        \end{align}
        Since this quantity is non-negative and vanishes identically for purely unitary gauge link configurations, it can be used as a measure of ``distance'' from the unitary subgroup. In other systems, small values of $F[U]$ have been empirically shown to be associated with correct convergence  \cite{Seiler:2012wz, Aarts:2013uxa}.
        
        Our numerical results for $F[U]$ as a function of Langevin time are shown in \fig\ref{fig:unorm}: starting from an initially unitary configuration in {SU($N_c$)}, the CL evolution drives the system away from unitarity into the non-compact directions of the complexified gauge group {SL($N_c$, $\mathbb C$)}. This leads to a growing $F[U]$ with increasing Langevin time. Without stabilization (left panels of \fig\ref{fig:unorm}), the unitarity norm grows quickly and plateaus at a large value after a short time. The particular Langevin time at which the plateau is reached approximately coincides with the time where the stochastic process has transitioned towards wrong convergence
        (compare with \fig\ref{fig:avg_spat_plaq_a}). 
        On the other hand, using gauge cooling and the anisotropic kernel, we observe a drastically slowed down growth of the unitarity norm (right panels of \fig\ref{fig:unorm}). Although the growth is already strongly reduced by gauge cooling alone, we find that our anisotropic kernel can reduce the increase of $F[U]$ over time even further. 
        Hence, our simulation results indicate that also the unitarity norm can be used as a validating observable for the metastable region, \new{at least on a qualitative level. For instance, the steady growth of the norm for $\tan \alpha = 0.5$ suggests that the metastable region will end eventually. Interestingly, for $\tan \alpha = 1$ the norm remains on the same level even at late Langevin times in our simulations. This could be an indication that in this case our simulations can be extended to later Langevin times without leaving the metastable region.} 
                
    \subsection{Systematics of the anisotropic kernel approach} \label{sec:systematics}

        \begin{figure} [!t]
            
                \centering\includegraphics[width=0.49\textwidth]{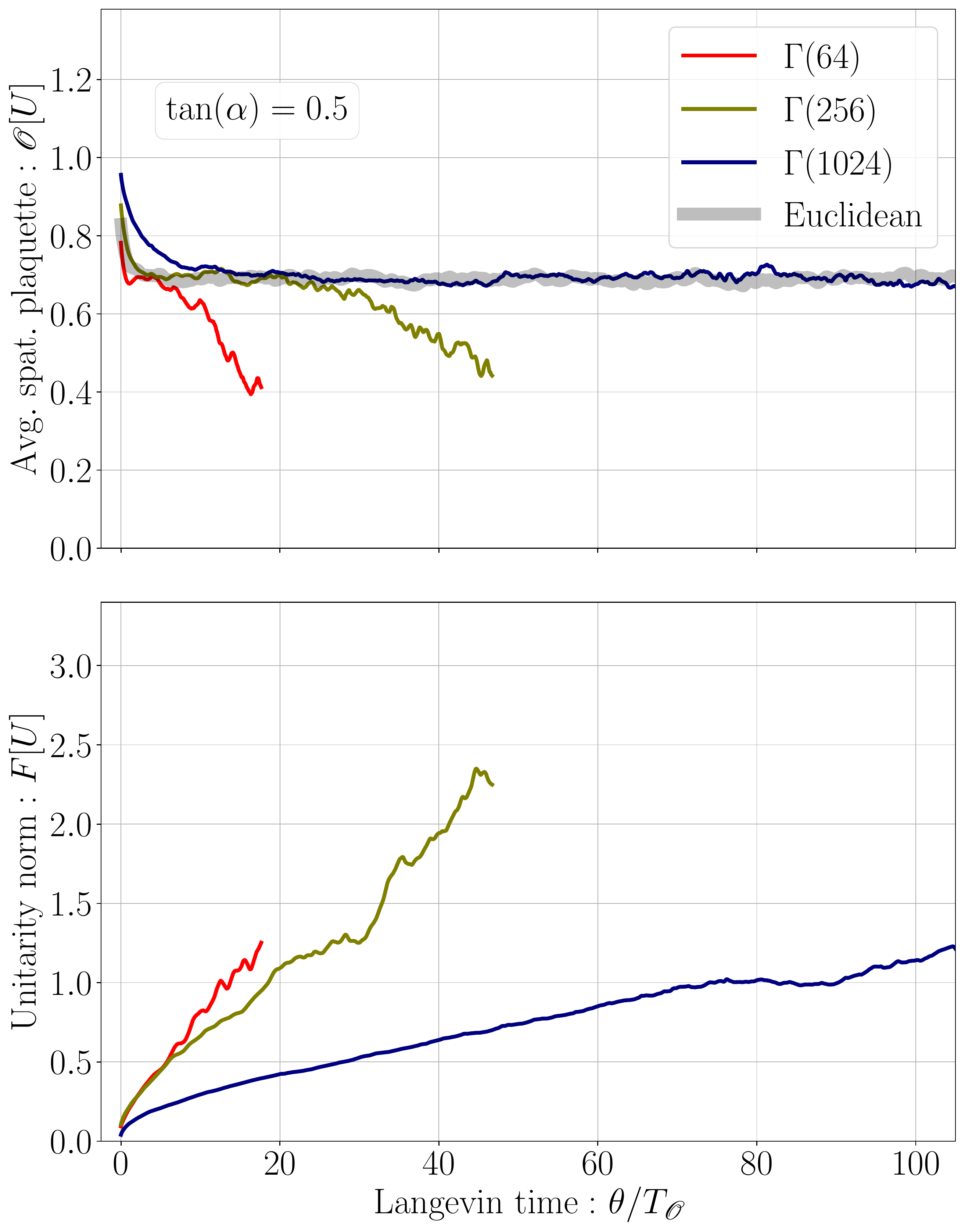}
                \centering\includegraphics[width=0.49\textwidth]{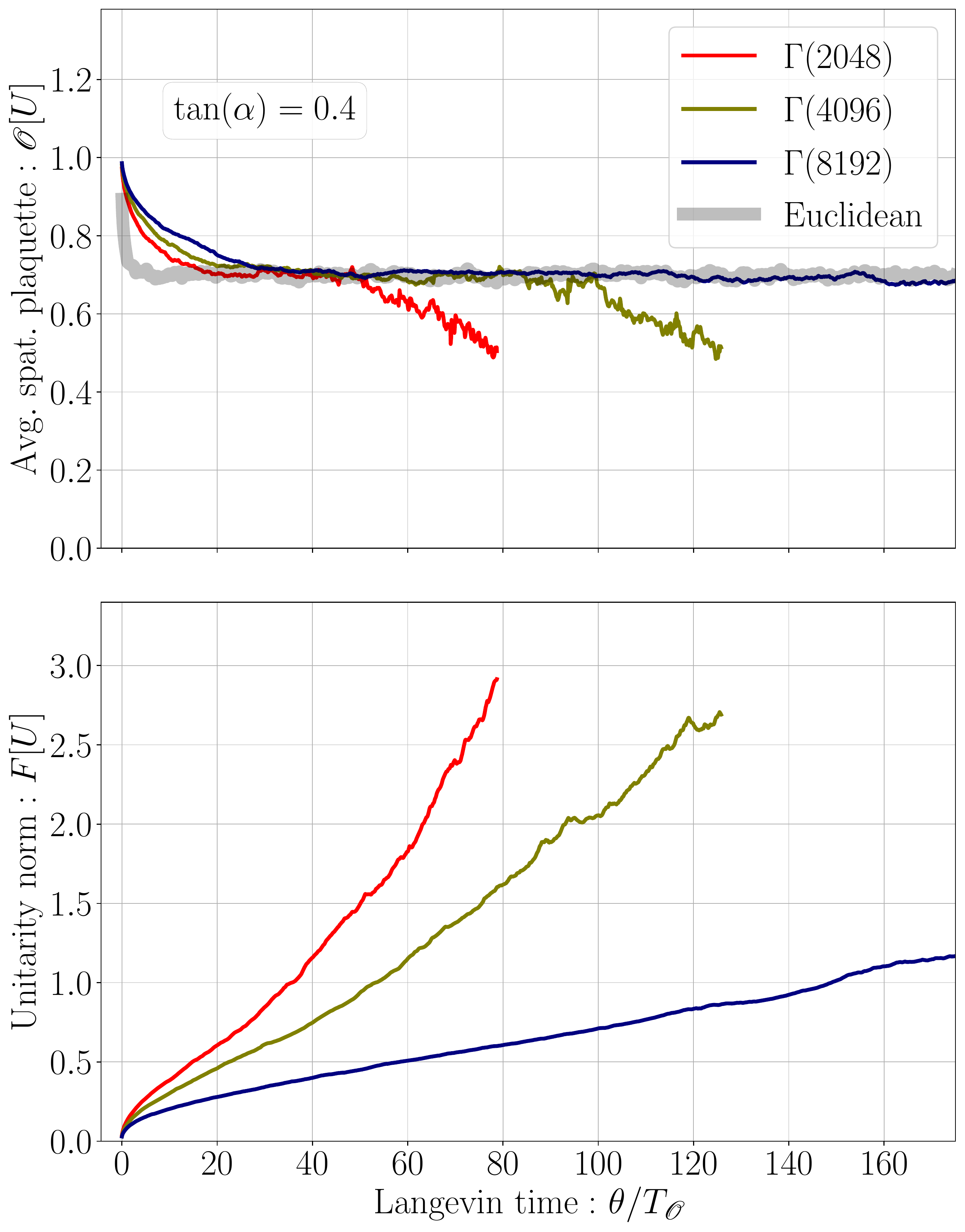}
            \caption{\label{fig:tan05_04_obs_unorm} 
             \new{Results for (\emph{top}) the spatial plaquette $\mathscr{O}$ in \mbox{Eq.~\eqref{eq:avg_spat_plaquette}} and (\emph{bottom}) the unitarity norm $F[U]$ for the tilt angles (\emph{left}) $\tan(\alpha)=0.5$ and (\emph{right}) $\tan(\alpha)=0.4$. We use AS, GC and our anisotropic kernel $\Gamma(N_t)$ for an increasing number of temporal lattice sites $N_t$ on an isosceles contour. Note that for $\tan(\alpha)=0.4$ the real-time extent is larger than the inverse temperature.}
            }
        \end{figure}

        \begin{figure} [t]
            \centering
                \includegraphics[width=0.49\textwidth]{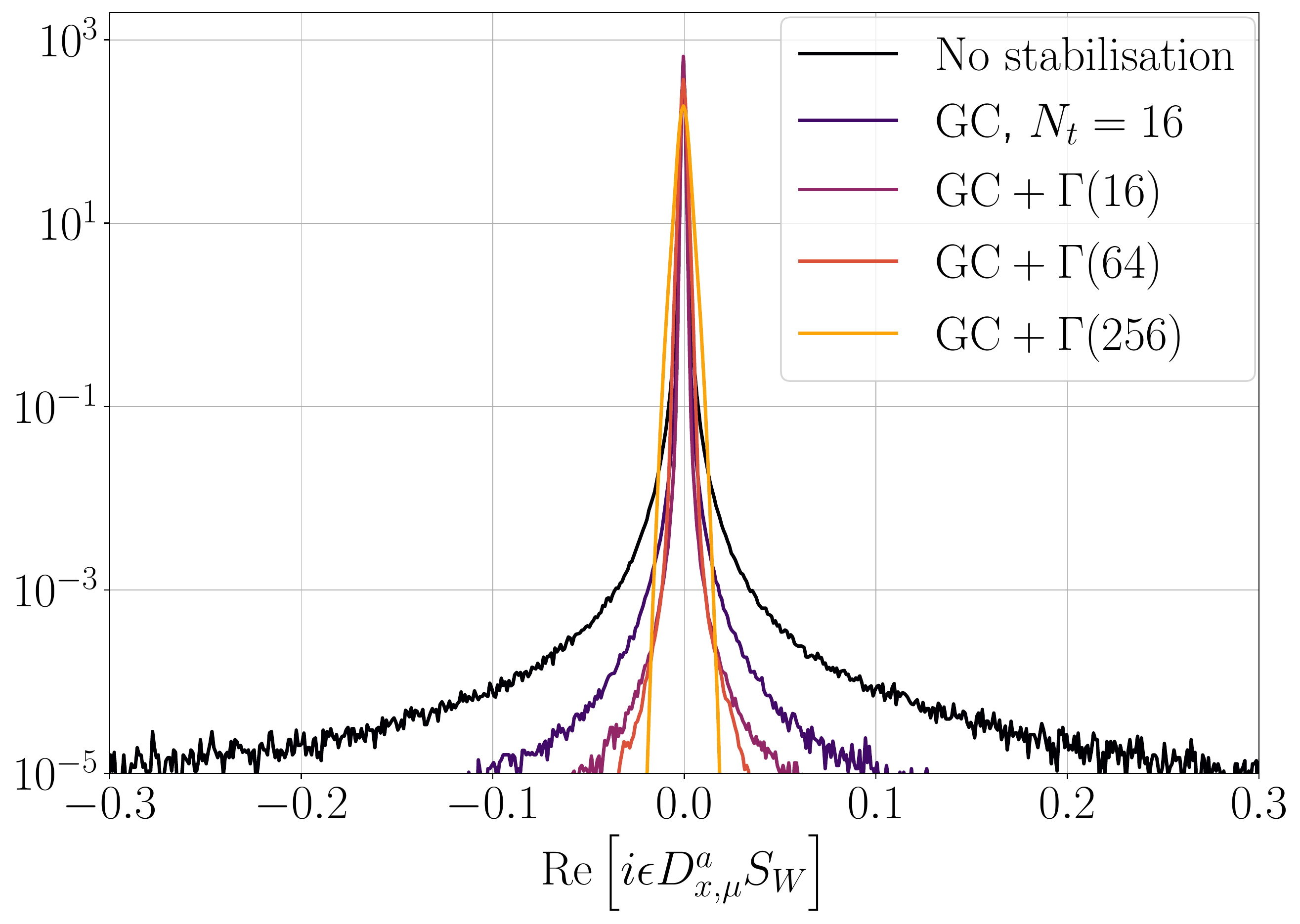}
                \includegraphics[width=0.49\textwidth]{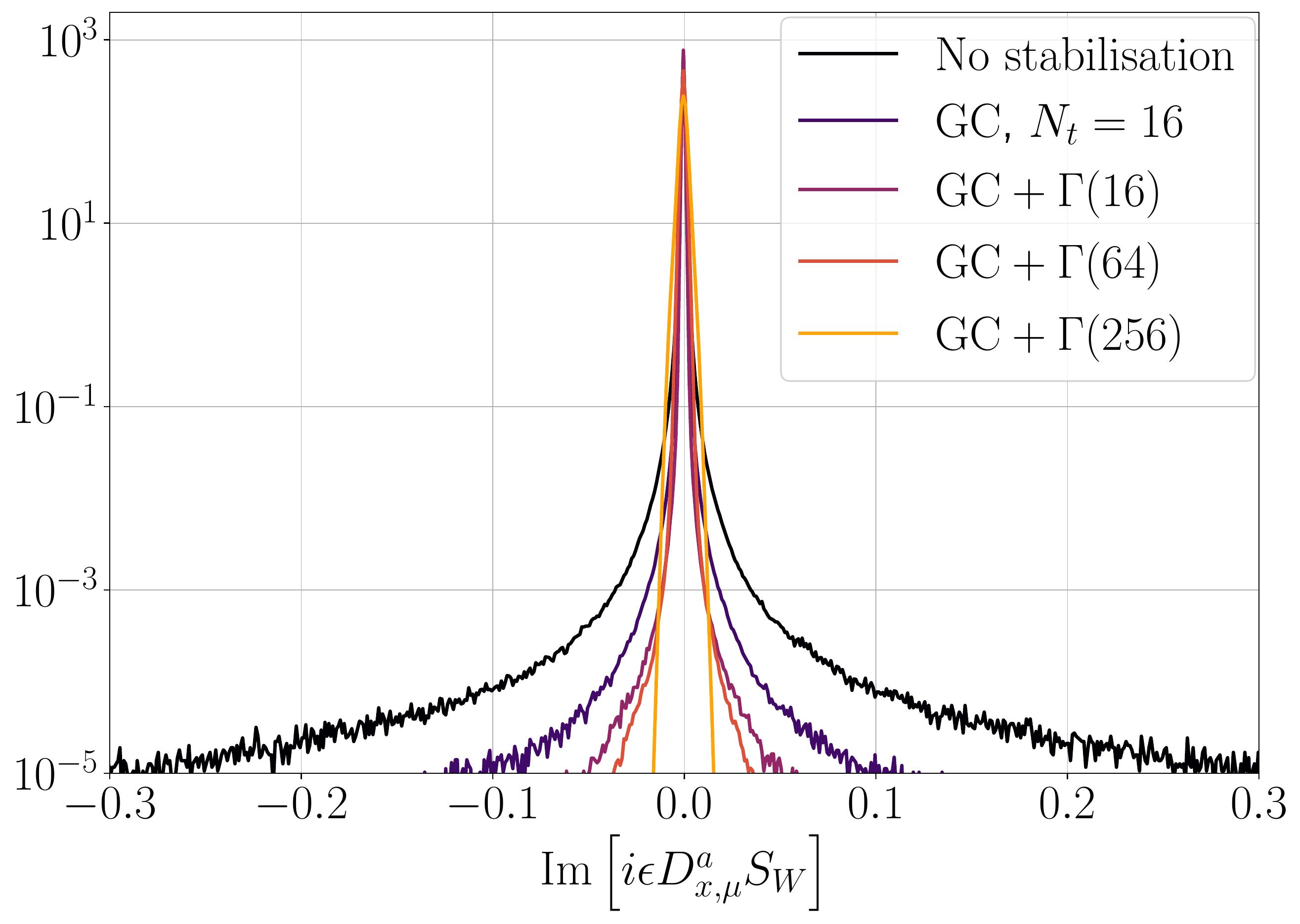}
            \caption{
            \new{Normalized histograms (distributions) of the real and imaginary parts of the drift term $D^a_{x,\mu} S_\mathrm{W} \equiv \delta S_\mathrm{W}/\delta \tilde A^a_{x,\mu}$ rescaled by $i\epsilon$ with $\epsilon = 10^{-4}$ for each degree of freedom without averaging over the lattice. The drift term was recorded in the metastable region of the simulations on an isosceles contour with tilt angle $\tan(\alpha)=0.625$ using the indicated stabilization techniques and number of temporal lattice sites $N_t$.
            }
            \label{fig:histogram}}
        \end{figure}
    
        The anisotropic kernel effectively increases the Langevin time spent in the metastable region with increasing number of lattice points along the time contour at fixed maximal real-time extent and inverse temperature $\beta$.
        To better understand this systematic behaviour, we perform multiple simulations at a fixed tilt angle, while varying the discretization of the time contour. The results are shown in the upper panels of \fig\ref{fig:tan05_04_obs_unorm} for the average spatial plaquette on an isosceles contour with $\tan(\alpha)=0.5$ and $\tan(\alpha)=0.4$. In the latter case, the real-time extent exceeds the inverse temperature $\beta$ (see \eq\eqref{eq:tilt_angle}). We observe that the width of the plateau, where the observable fluctuates around the correct value, increases relative to the autocorrelation time $T_\mathscr{O}$ for increasing $N_t$. Simulations using the anisotropic kernel are, however, not indefinitely stable. We can see that the stochastic process deviates from the correct values at large Langevin times. Nevertheless, it is apparent that the extent of the metastable region can be systematically enlarged in order to calculate expectation values to sufficient accuracy as shown in \tab\ref{tab:avg_spat_plaq}. We emphasize that this systematic behaviour of our kerneled CL update step is also seen in the evolution of the unitarity norm depicted in the lower panels of \fig\ref{fig:tan05_04_obs_unorm}. With increasing $N_t$ its values remain small for a longer Langevin time region, which indicates prolonged stability of the CL simulations as discussed in \se\ref{sec:monitoring}. 
        Consequently, our approach allows us -- for the first time --  to simulate non-Abelian gauge theories with larger real-time extent than the inverse temperature, albeit in a metastable manner.%
        \footnote{
        Note that this approach typically requires more computational resources for decreasing tilt angles as we need finer discretizations to reach large enough plateaus of the metastable region. As the kernel is controlled by adapting the temporal lattice spacing, the memory requirements only grow linearly with $N_t$. Computational time grows more quickly, since the autocorrelation time also grows with $N_t$. Therefore, more update steps are required on a larger lattice to sample enough uncorrelated data points. 
        }
        
        \new{We emphasize that here we go to large $N_t$ and thus anisotropic lattices in order to demonstrate the systematic growth of the metastable region. However, for practical purposes it will be sufficient to use discretizations with smaller $N_t$ such as $N_t = 256$ or even smaller for $\tan \alpha = 0.5$ in \mbox{\fig\ref{fig:tan05_04_obs_unorm}}. In this case sufficient statistics to compute observables can be achieved by performing multiple independent simulations in addition to averaging over the metastable region. This also reduces artifacts that stem from the anisotropic lattice spacings and simplifies a proper renormalization procedure, which we will report elsewhere.}

        \new{
        Another important aspect of the convergence of CL towards the correct probability density function is indicated by a sufficiently fast decay of the drift $D^a_{x,\mu} S_\mathrm{W} \equiv \delta S_\mathrm{W}/\delta \tilde A^a_{x,\mu}$ that enters the CL equation \mbox{\cite{Nagata:2016vkn, Aarts:2009uq}}. Intuitively, the histograms show how fast the stochastic process strays into the complex configuration space, which leads to instabilities or wrong convergence. Figure \mbox{\ref{fig:histogram}} shows the normalized histograms of the real and imaginary parts of the drift term for the simulation of an isosceles contour with tilt angle $\tan(\alpha)=0.625$. The values are recorded during the metastable region without averaging over the lattice. We observe that our anisotropic kernel gradually contracts the histogram with growing $N_t$. We emphasize that this behavior cannot be observed by increasing $N_t$ without the use of our kernel and the GC method alone only slightly narrows the histograms for the simulated system. Our kernel leads to localized histograms for $N_t=256$ without any skirts. This suggests that the criterion of correctness of CL, that is discussed in great detail in \re\mbox{\cite{Nagata:2018net}}, may be satisfied by the introduction of our kernel for the observed system.
        } 

\section{Conclusion}
    \label{sec:conclusion}
        
    In this work, we have revisited the CL method for non-Abelian gauge theories on complex time paths, as required for simulations in physical time. We have stated a time-reversal symmetric and unambiguous formulation of CL equations both in the continuum and numerically on the lattice. We found that the introduction of an anisotropic kernel allows for systematic improvement of the stability of our simulations. 
    
    In particular, our first objective was to obtain a consistent formulation of the CL equation. We have resolved the ambiguities of the traditional CL equation for a simple quantum mechanical toy model which we then generalized to Yang-Mills theory on complex time contours. 
    The main issue is that commonly used CL equations are ambiguous on complex time paths due to Dirac delta distributions of the time coordinates, which appear in the noise correlators. Therefore, we introduced a contour parameter formulation of the CL equation. Exploiting the kernel freedom, we have shown that it is parametrization independent and consistent with the known CL equations for time paths along the real and imaginary axis, respectively. For numerical simulations, the obtained CL equations were discretized in a time-symmetric manner such that they correctly approach the corresponding continuum limit for small lattice spacings. 
    We additionally exploited the kernel freedom for Yang-Mills theory to obtain a novel CL update scheme given by \eqs\eqref{eq:new_cle_temp} and \eqref{eq:new_cle_spat} that differs from the traditional equations in terms of an anisotropic kernel. 
    
    We have then demonstrated for an SU(2) gauge theory in 3+1 dimensions that our new CL equations lead to remarkable improvements regarding stability and convergence.
    Previously, the main problem has been that CL simulations with earlier update equations suffered from severe instabilities and converged to wrong results. In particular, these issues become significant for simulations of Yang-Mills theory on tilted time contours, which, in the limit of vanishing tilt angle, approximate the Schwinger-Keldysh time contour required for Yang-Mills theory in physical time. 
    We have shown here that recently developed stabilization techniques such as adaptive step size, gauge cooling and dynamical stabilization mitigate these problems but are insufficient when the tilt angle of the discretized time path becomes too small. Dynamical stabilization even introduces a bias that becomes highly problematic at small tilt angles. 
    Our new CL update scheme paves the way towards resolving these issues. Even in the case of small tilt angles, the CL evolution exhibits a metastable region of correct convergence, whose lifetime can be prolonged systematically by increasing the resolution along the complex time contour.
    
    Using the expectation value of the average spatial plaquette as an example, 
    we have demonstrated that we obtain correct results for any of the previously studied tilt angles by taking samples from this region. We validated our results in multiple ways: Exploiting the time invariance of the observable, we compared our results with stable Langevin simulations along the Euclidean time contour. Moreover, we have studied the unitarity norm, Dyson-Schwinger equations, \new{and the distribution of the drift term entering the CL update step}. We have found that in the metastable region, the unitarity norm remains small and that both sides of the Dyson-Schwinger equations coincide with high accuracy as opposed to regions of wrong convergence which exhibit much larger fluctuations. \new{Our anisotropic kernel additionally narrows the distribution of the drift term, which stabilizes the CL evolution.} 
    This procedure can be extended systematically to larger metastable regions or to smaller angles. For the first time with Yang-Mills simulations, this enabled us to obtain correct results on a tilted time contour whose real-time extent exceeds its inverse temperature. 
    
    Our novel approach introduces a powerful and promising tool towards computing real-time observables and non-equilibrium dynamics of gauge theories in CL on a Schwinger-Keldysh time contour.
    In perspective, this may allow us to calculate transport coefficients, viscosities and spectral functions in QCD directly from first principles, which have important applications in heavy-ion collisions and beyond. Conceptually, our method is indeed able to yield results for unprecedentedly small tilt angles, potentially allowing us to perform the limit to a Schwinger-Keldysh time contour. However, this requires large lattices and high resolution in the temporal direction. We will investigate the possibilities of our framework by extending our analysis to additional observables, 
    the assessment of boundary terms, the extraction of unequal-time correlation functions and the approach towards a continuous Schwinger-Keldysh time contour in forthcoming studies.
    \new{Furthermore, the extraction of physical observables from lattice simulations will also necessitate the determination of the physical lattice spacing and the renormalized lattice anisotropy.} 
    \new{Beyond real-time applications, }the revision of the discretized CL equation in this work and the introduction of a novel anisotropic kernel may also benefit other applications of the CL method, in particular QCD at finite chemical potential. We intend to investigate this exciting direction in the future.

\begin{acknowledgments}
  The authors would like to thank D.~Alvestad and D.~Sexty for valuable discussions regarding the basic foundations of complex Langevin and rescalings as part of the kernel freedom. We are further grateful to J.M.~Pawlowski, A.~Rebhan and F.P.G.~Ziegler for very useful discussions and comments, and to A.~Ipp for technical input regarding code development.
  This research was funded by the Austrian Science Fund (FWF) project P~34455-N. Moreover, Paul Hotzy expresses his gratitude to the Doktoratskolleg Particles and Interactions (DK-PI,  FWF doctoral program No.~W-1252-N27).
  The computational results presented have been achieved in part using the Vienna Scientific Cluster (VSC).
\end{acknowledgments}

    
\appendix

\section{Discretized CL equation on a complex time contour} 
    \label{app:disc_drift}

        For numerical simulations we need a discretized formulation of the CL equation. Using the time-reversal symmetric discretized action $S_{\mathrm{latt}}$ in \eqref{eq:lambda_action_discr}, that we state here again for completeness,
        \begin{align} 
            S_{\mathrm{latt}} [x] = \sum_{k=1}^{N_t} (\lambda_{k} - \lambda_{k-1}) \frac{t_{k} - t_{k-1}}{\lambda_{k} - \lambda_{k-1}} \left( \frac{1}{2} \left( \frac{x_{k} - x_{k-1}}{t_{k} - t_{k-1}} \right)^2 - \frac{1}{2} \left(V(x_{k}) + V(x_{k-1}) \right)
            \right),
        \end{align}
        its variation reads
        \begin{align}
            \delta S_{\mathrm{latt}} [x] = \sum_k  
            \left( - \left( \frac{x_{k+1} - x_k}{a_{t,k}} - \frac{x_k - x_{k-1}}{a_{t,k-1}} \right) - \frac{1}{2}(a_{t,k} + a_{t,k-1}) V'(x_k) \right) \delta x_k,
        \end{align}
        where the $\lambda$-spacings are $a_{\lambda,k} = \lambda_{k+1} - \lambda_k$ and the time steps are $a_{t,k} = t_{k+1} - t_k$.
        
        In order to define a quantity that approaches the functional derivative in the limit of infinitesimal time steps, we write the discrete variation as
        \begin{align}
            \delta S_{\mathrm{latt}} [x] = \sum_k \frac{1}{2} (a_{\lambda,k}  + a_{\lambda,k-1})\, \frac{\delta S_{\mathrm{latt}}}{\delta x(\lambda_k)} \delta x_k,
        \end{align}
        which is the discrete analogue of \eq \eqref{eq:lambda_variation}. This allows us to read off the discrete functional derivative
        \begin{align} \label{eq:discrete_functional_derivative}
             \frac{\delta S_{\mathrm{latt}}}{\delta x(\lambda_k)}  = 
            \frac{1}{\frac{1}{2}(a_{\lambda,k} + a_{\lambda,k-1})} 
            \left( - \frac{x_{k+1} - x_k}{a_{t,k}} + \frac{x_k - x_{k-1}}{a_{t,k-1}}  -  \frac{1}{2}(a_{t,k} + a_{t,k-1}) V'(x_k) \right),
        \end{align}
        which in the limit of $a_{\lambda,k} \rightarrow 0$ yields the correct result of \eq \eqref{eq:lambda_drift}. Note that the discrete functional derivative is related to the usual derivative of $S_\mathrm{latt}$ via
        \begin{align}
            \frac{\delta S_{\mathrm{latt}}}{\delta x(\lambda_k)} = \frac{1}{\frac{1}{2}(a_{\lambda,k} + a_{\lambda,k-1})} \frac{\partial S_\mathrm{latt}}{\partial x_k}.
        \end{align}

        A natural choice for the parameter of the time path is the arc length. This amounts to choosing
        \begin{align}
            a_{\lambda,k}  = \vert a_{t,k} \vert,
        \end{align}
        for all $k$ and yields
        \begin{align}
            \frac{\delta S_\mathrm{latt}}{\delta x(\lambda_k)} = 
            \frac{1}{\frac{1}{2}(\vert a_{t,k} \vert + \vert a_{t,k-1} \vert)} 
            \left( - \frac{x_{k+1} - x_k}{a_{t,k}} + \frac{x_k - x_{k-1}}{a_{t,k-1}}  -  \frac{1}{2}(a_{t,k} + a_{t,k-1}) V'(x_k) \right).
        \end{align}
        We proceed by discretizing the noise term of the CL equation along the contour. For a small spacing of the contour parameter $a_{\lambda,k}  \to 0$, we may approximate the two-point correlation function in the following way:
        \begin{align}
        \begin{split}
            \langle \eta(\theta', \lambda_{k'}) \eta(\theta, \lambda_k) \rangle
            &= 2 \delta(\theta - \theta')\, \frac{1}{\frac{1}{2}(a_{\lambda,k} + a_{\lambda,k-1})}\, \delta_{k k'}  \\
            &= 2 \delta(\theta - \theta')\, \frac{1}{\frac{1}{2}(|a_{t,k}| + |a_{t,k-1}|)} \delta_{k k'} \\
            &= \frac{1}{\frac{1}{2}(|a_{t,k}| + |a_{t,k-1}|)} \langle \eta_{k'}(\theta') \eta_{k}(\theta) \rangle,
        \end{split}
        \end{align}
        where $\delta_{kk'}$ is the Kronecker delta and the discrete (with respect to $\lambda$) noise correlator reads
        \begin{align}
            \langle \eta_{k'}(\theta') \eta_{k}(\theta) \rangle = 2 \delta(\theta - \theta')  \delta_{k k'}.
        \end{align}
    The factor $(|a_{t,k}| + |a_{t,k-1}|) / 2$ corresponds to the averaged time step centered around $t_k$.
    
    We can now formulate the CL equation with a discretized contour parameter for the considered toy model as
        \begin{align} \label{eq:lambda_cle}
            \frac{d z_k(\theta)}{d \theta} = 
            i \frac{2}{|a_{t,k}| + |a_{t,k-1}|} \left. \frac{\partial S_\mathrm{latt}}{\partial z_k} \right\vert_{\theta} + \sqrt{\frac{2}{|a_{t,k}| + |a_{t,k-1}|}}\, \eta_k(\theta).
        \end{align}
    Finally, we discretize the Langevin time $\theta$ in steps of $\epsilon$ which yields the Euler-Maruyama scheme 
    \begin{align} \label{eq:lambda_cle_disc_app}
        z_k(\theta + \epsilon) = z_k(\theta) + 
        i \frac{\epsilon}{\frac{1}{2}(|a_{t,k}| + |a_{t,k-1}|)} \left. \frac{\partial S_\mathrm{latt}}{\partial z_k} \right\vert_{\theta}  +
        \sqrt{\frac{\epsilon}{\frac{1}{2}(|a_{t,k}| + |a_{t,k-1}|}}\, \tilde \eta_k(\theta).
    \end{align}
    The correlator of the discrete noise is given by
    \begin{align}
        \langle \tilde \eta_{k'}(\theta') \tilde \eta_k(\theta) \rangle = 2 \delta_{\theta \theta'} \delta_{kk'}.
    \end{align}

\section{Parameterizing the time contour for gauge theories} 
    \label{app:ym_contour_param}
        
        In this Appendix we provide the details of our considerations in \se \ref{sec:ym_contour_param} for gauge theories in a contour parameter formulation. Let us recall \eq \eqref{eq:lambda_ym_cle}:
        \begin{align} \label{eq:lambda_ym_cle_app}
        \begin{split}
            \p_\theta A^a_\mu(\theta, \lambda, \mathbf x) &= i \left. \frac{\dd S_\mathrm{YM}}{\dd A^a_\mu(\lambda, \mathbf x)} \right\vert_{\theta} + \eta^a_\mu(\theta, \lambda, \mathbf  x), \\
            \langle \eta^a_\mu(\theta, \lambda, \mathbf  x) \rangle &= 0, \\
            \langle \eta^a_\mu(\theta, \lambda, \mathbf x) \eta^b_\nu(\theta', \lambda', \mathbf x') \rangle &= 2 \delta_{\mu\nu} \delta^{ab} \delta(\theta - \theta') \delta(\lambda - \lambda') \delta^{(3)}(\mathbf x- \mathbf x')\,,
        \end{split}
        \end{align}
        with $\mu, \nu \in \{ \lambda, x, y, z\}$.
        Note that it involves an unambiguous Dirac delta distribution $\delta(\lambda - \lambda')$ and is written in terms of the $\lambda$-component $A^a_\lambda(\lambda, \mx)$.
        
        The Yang-Mills action can be written using an integral over the contour parameter
        \begin{align}
            S_\mathrm{YM} = - \frac{1}{4}  \int d\lambda d^3x\, \frac{dt}{d\lambda} F^{\mu \nu}_a F_{\mu \nu}^a.
        \end{align}
        We view the parameterization $t(\lambda)$ as a change of the time coordinate $t(\lambda) \rightarrow \lambda$. As such, the contraction of Lorentz indices must be carried out with the appropriate metric
        \begin{align}
            ds^2 = g_{\mu\nu} dx^\mu dx^\nu = \left(\frac{dt}{d\lambda} \right)^2 d\lambda^2 - dx^i dx^i,
        \end{align}
        and the components of the gauge field transform according to
        \begin{align} \label{eq:lambda_field_transformation}
            A^a_\lambda(\lambda, \mx) = \frac{dt}{d\lambda} A^a_t(t, \mx) , \qquad A^a_i(\lambda, \mx) = A^a_i(t, \mx).
        \end{align}
        The functional derivative of the action can be rewritten in terms of $t$-derivatives. A comparison of the integrands of the variation of $S_\mathrm{YM}$ 
        \begin{align}
            \delta S_\mathrm{YM} = \int dt d^3x \, \frac{\delta S_\mathrm{YM}}{\delta A^a_\mu(t,\mx)} \delta A^a_\mu(t, \mx) = \int d\lambda d^3x \, \frac{\delta S_\mathrm{YM}}{\delta A^a_\mu(\lambda, \mx)} \delta A^a_\mu(\lambda, \mx)
        \end{align}
        yields the relations 
        \begin{align}
            \frac{\delta S_\mathrm{YM}}{\delta A^a_t(t,\mx)} = \frac{\delta S_\mathrm{YM}}{\delta A^a_\lambda(\lambda,\mx)}, \qquad  \frac{dt}{d\lambda} \, \frac{\delta S_\mathrm{YM}}{\delta A^a_i(t,\mx)} = \frac{\delta S_\mathrm{YM}}{\delta A^a_i(\lambda,\mx)}. 
        \end{align}
        We can therefore write \eq\eqref{eq:lambda_ym_cle_app} in terms of the functional derivative with respect to fields in physical time (stated as \eq \eqref{eq:cle_ym_contour_hybrid} in the main text)
        \begin{align} \label{eq:cle_ym_contour_hybrid_app}
        \begin{split}
            \p_\theta A^a_t(\theta, t, \mx) &= i \frac{d\lambda}{dt}\left.\frac{\dd S_\mathrm{YM}}{\dd A^a_t(t, \mx)}\right\vert_{\theta} + \frac{d\lambda}{dt}\, \eta^a_\lambda(\theta, \lambda, \mx), \\
            \p_\theta A^a_i(\theta, t, \mx) &= i \frac{dt}{d\lambda}\left.\frac{\dd S_\mathrm{YM}}{\dd A^a_i(t, \mx)}\right\vert_{\theta} + \eta^a_i(\theta, \lambda, \mx).
        \end{split}
        \end{align}

        In analogy to \se\ref{sec:param_contour}, we now consider a Minkowski time contour $t(\lambda) \in \mathbb R$ to show that we can reproduce \eq\eqref{eq:ym_cle_mink} with our contour parameter CL equation \eqref{eq:lambda_ym_cle_app}. For this we can relate the noise correlator $\langle \eta^a_\mu(\theta, \lambda, \mathbf x) \eta^b_\nu(\theta', \lambda', \mathbf x') \rangle$ 
        to an expression using a Dirac distribution for real-valued time arguments in an unambiguous manner. This naturally leads to the transformation behaviour of the noise field
        \begin{align}
            \eta^a_\lambda(\theta, \lambda, \mx)=\sqrt{\frac{dt}{d\lambda}}\, \eta^a_t(\theta, t, \mx), \qquad \eta^a_i(\theta, \lambda, \mx)=\sqrt{\frac{dt}{d\lambda}}\, \eta^a_i(\theta, t, \mx),
        \end{align}
        which yields the original correlator
        \begin{align}
            \langle \eta^a_\mu(\theta, t, \mx) \eta^b_\nu(\theta', t', \mx') \rangle = 2 \delta_{\mu\nu} \delta^{ab} \delta(\theta-\theta') \delta(t-t') \delta^{(3)}(\mx - \mx').
        \end{align}
        Note that even though the spatial component of the gauge field does not acquire a factor when transforming from $\lambda$ to $t$ (see \eq\eqref{eq:lambda_field_transformation}), the spatial noise field $\eta_i(\theta, t, \mx)$ must be multiplied with $\sqrt{dt/d\lambda}$ to reproduce the correlator in the Minkowski case. Taking the Minkowski action $S_M$ into account, we can now put the pieces together and obtain 
        \begin{align}\label{eq:mink_kerneled_cle}
            \begin{split}
            \p_\theta A^a_t(\theta, t, \mx) &= i \,\frac{d\lambda}{dt}\left.\frac{\dd S_\mathrm{M}}{\dd A^a_t(t, \mx)}\right\vert_{\theta} + \sqrt{\frac{d\lambda}{dt}}\,\eta^a_t(\theta, t, \mx), \\
            \p_\theta A^a_i(\theta, t, \mx) &= i \,\frac{dt}{d\lambda}\left.\frac{\dd S_\mathrm{M}}{\dd A^a_i(t, \mx)}\right\vert_{\theta} + \sqrt{\frac{dt}{d\lambda}}\,\eta^a_i(\theta, t, \mx). 
            \end{split}
        \end{align}

        We observe that for parametrized Minkowski time contours, the CL equation in \eq\eqref{eq:mink_kerneled_cle} represents a kerneled version of the original Minkowski formulation in \eq\eqref{eq:ym_cle_mink}. The kernel which relates these equations reads
        \begin{align}
            \Gamma^{ab}_{\mu\nu}(t, x, t', x') = \delta(t-t')\delta^{(3)}(\mx-\mx') \delta^{ab} \left(\frac{d\lambda}{dt} \delta_{\mu t} \delta_{\nu t} + \frac{dt}{d\lambda} \delta_{\mu i}\delta_{\nu i}\right).
        \end{align}
        Therefore, the parameterized version of the Minkowski CL equation retains the stationary solution of the formulation of CL for Minkowski time contours in \eq\eqref{eq:ym_cle_mink}. The same can be shown for the Euclidean CL equation \eqref{eq:ym_cle_eucl}. 

    \section{Details on discretizing the CL equation for Yang-Mills theory} 
    \label{app:lattice}
        
        Here we provide details on the derivation of the discrete CL equations in \se \ref{sec:lattice}
        using the following steps:
        \begin{enumerate}
        \item We first discretize the Yang-Mills action $S_\mathrm{YM}[A_\mu]$ in terms of gauge links $U_{x,\mu}$, yielding the quadratically accurate Wilson action $S_\mathrm{W}[U]$ for a discretized complex time contour $t(\lambda_k)$.
        \item We then relate the drift term $\delta S_\mathrm{YM} / \delta A^a_\mu(t,x)$ appearing in the continuum CL equation to the group derivative $D^a_{x,\mu} S_\mathrm{W}[U]$ of the Wilson action.
        \item Finally, putting the pieces together, we approximate the continuous CL equations in \eq\eqref{eq:cle_ym_contour_hybrid} for gauge fields as discrete CL equations for gauge links in \eqs\eqref{eq:ym_contour_cle_unkerneled_temp} and \eqref{eq:ym_contour_cle_unkerneled_spat}, which correctly approach the continuum limit for small lattice spacings.
        \end{enumerate}
        
        \subsection{Approximating the Yang-Mills action}
        A general Wilson line along an arbitrary path%
        \footnote{Not to be confused with the complex time contour $\mathscr C$.} 
        $\mathcal C$ is given by
        \begin{align}
                U_\mathcal{C}= {\mathcal P}\exp \left( i g \intop_0^1 ds \frac{dx^\mu(s)}{ds} A_\mu(x(s))  \right) \in \mathrm{SL}(N_c, \mathbb C),
        \end{align}
        with $A_\mu = A^a_\mu t^a$. The path $\mathcal C$ is parameterized by $x^\mu(s)$ with $s\in[0,1]$ with startpoint $x^\mu(0)$ and endpoint $x^\mu(1)$, and $t^a$ are the traceless Hermitian generators of SU($N_c$). The symbol ${\mathcal P}$ denotes path ordering defined by
        \begin{align}
            {\mathcal{P}} \left[ X(s_1) X(s_2) \right] = \begin{cases}
            X(s_1) X(s_2), \quad s_1 < s_2, \\
            X(s_2) X(s_1), \quad s_1 \geq s_2.
            \end{cases}
        \end{align}
        Under a general gauge transformation $\Omega(x) \in \mathrm{SL}(N_c, \mathbb C)$
        \begin{align}
            A'_\mu(x) = \Omega(x) \left( A_\mu(x) + \frac{1}{ig} \partial_\mu \right) \Omega^{-1}(x),
        \end{align}
        the Wilson line transforms according to
        \begin{align}
            U'_\mathcal{C} = \Omega(x(0)) \, U_\mathcal{C} \, \Omega^{-1}(x(1)).
        \end{align}
        In the limit of small lattice spacings, gauge links can be approximated by matrix exponentials of the gauge fields evaluated at the mid-points of lattice edges
        \begin{align}
            U_{x,t} &\approx \exp \left( i g a_{t,k} \,A_t\left(t_k + \frac{1}{2} a_{t,k}, \mathbf x\right) \right), \\
            U_{x,i} &\approx \exp \left( i g a_s \,A_i\left(t_k , \mathbf x + \frac{1}{2} a_s \hat e_i\right) \right),
        \end{align}
        where subscript $x$ is a shorthand for the lattice site at time $t_k$ and position $\mx$. More compactly, we write        
        \begin{align} \label{eq:link_continuum}
            U_{x,\mu} &\approx \exp \left( i g a_\mu A_\mu\left(x+\frac{1}{2} \hat \mu\right) \right),
        \end{align}
        where no sum over $\mu$ is implied. Links with a negative index $U_{x,-\mu}$ point into the opposite direction, i.e.~$U_{x,-\mu} = U_{x-\mu, \mu}^{-1}$.
        Plaquettes $U_{x,\mu\nu}$ are defined as $1 \times 1$ Wilson loops
        \begin{align}
            U_{x,\mu\nu} = U_{x,\mu} U_{x+\hat\mu, \nu} U_{x+\hat\nu, \mu}^{-1} U_{x,\nu}^{-1},
        \end{align}
        which, in the limit of small lattice spacings, approximate the field strength tensor at the mid-point of the face spanned by directions $\hat\mu$ and $\hat\nu$
        \begin{align}
            U_{x,\mu\nu} \approx \exp \left( i g a_\mu a_\nu \,F_{\mu\nu}\left(x+\frac{1}{2} \hat\mu + \frac{1}{2} \hat \nu \right) \right).
        \end{align}
        Using the above approximation, we discretize the Yang-Mills action in \eq\eqref{eq:ym_action} in terms of plaquettes, which yields the Wilson action
        \begin{align} \label{eq:wilson_action}
            S_\mathrm{W}[U] &= \sum_{k, \mx} \bigg( -\frac{2N_c}{g^2} \frac{a_s}{a_{t,k}} \sum^3_{i=1} \frac{1}{2N_c} \mathrm{Tr} \left[ U_{x,0i} + U^{-1}_{x,0i} - 2 \right] \nonumber \\
            &\qquad\quad + \frac{2N_c}{g^2} \frac{\bar a_{t,k}}{a_s} \sum^3_{i=1}\sum_{j\neq i} \frac{1}{4N_c} \mathrm{Tr} \left[ U_{x,ij} + U^{-1}_{x,ij} - 2 \right] \bigg).
        \end{align}
        Here we have used the averaged time-step 
        \begin{align}
            \label{eq:avg_timestep}
            \bar a_{t,k} = \frac{1}{2}\left(a_{t,k} + a_{t,k-1}\right) = \frac{1}{2}\left(t_{k+1} - t_{k-1}\right)
        \end{align}
        in the spatial plaquette term. This guarantees time reversal symmetry and quadratic accuracy of the Wilson action even for general time contours. 
        Adopting a similar notation to \cite{Berges:2006xc}, the Wilson action can be written more compactly as
        \begin{align} \label{eq:wilson_action_compact}
            S_\mathrm{W}[U] = \frac{1}{2N_c} \sum_{x, \mu \neq \nu} \beta_{x,\mu\nu} \mathrm{Tr}\left[U_{x,\mu\nu} - 1\right],
        \end{align}
        where we introduce the coupling constants
        \begin{align}
        \beta_{x,0i} &=\beta_{x,i0}=-\beta_{k,0}=-\frac{2N_c}{g^2}\frac{a_s}{a_{t,k}}, \\
        \beta_{x,ij}&=\beta_{x,ji}=+\beta_{k,s}=+\frac{2N_c}{g^2}\frac{\bar a_{t,k}}{a_s}.
        \end{align}
        As before, the index $k$ refers to the time slice index associated with the lattice site $x$.

        \subsection{Relating the drift terms}
        
        The drift terms appearing in the continuum CL evolution in \eq\eqref{eq:cle_ym_contour_hybrid} are the functional derivatives of the Yang-Mills action $S_\mathrm{YM}[A_\mu]$, which are related to its variation $\delta S_\mathrm{YM}$ via
        \begin{align} \label{eq:ym_variation}
            \delta S_\mathrm{YM}[A_\mu, \delta A_\mu] = \int dt \, d^3 x \,  \frac{\delta S_\mathrm{YM}}{\delta A^a_\mu(t, \mx)} \delta A^a_\mu(t, \mx). 
        \end{align}
        On the other hand, the Wilson action $S_\mathrm{W}[U]$ is a function of the group-valued gauge links $U_{x,\mu}$. A variation within the group SL($N_c$, $\mathbb C$) can be performed by varying the gauge links $U_{x,\mu} \rightarrow U_{x,\mu} + \delta U_{x,\mu}$ with
        \begin{align}
            \delta U_{x,\mu} = i  \delta \tilde A^a_{x, \mu} t^a U_{x,\mu}, \quad \delta U_{x,\mu}^{-1} = -i  \delta \tilde A^a_{x, \mu} U^{-1}_{x,\mu} t^a ,
        \end{align}
        where the dimensionless variation 
        \begin{align}
            \delta \tilde A^a_{x,\mu} = g a_\mu \delta A^a_{x,\mu} \in \mathbb C
        \end{align}
        is defined at the mid-point $x+ \hat\mu/2$. Expanding the Wilson action $S_\mathrm{W}[U+\delta U]$ to linear order in $\delta \tilde A^a_{x,\mu}$ yields
        \begin{align} \label{eq:wilson_variation}
            \delta S_\mathrm{W} = \sum_{x,\mu} \frac{\delta S_\mathrm{W}}{\delta \tilde A^a_{x,\mu}} \delta \tilde A^a_{x,\mu}\,.
        \end{align}
        The variation of the action thus reads
        \begin{align}
            \delta S_\mathrm{W}[U, \delta U] = \frac{1}{2N_c} \sum_{x,\mu\neq \nu} \beta_{x,\mu\nu} \mathrm{Tr} \left[ \delta U_{x,\mu\nu} \right],
        \end{align}
        where the variation of the plaquette is given by
        \begin{align}
            \delta U_{x,\mu\nu} &= \delta U_{x,\mu} U_{x+\mu, \nu} U^{-1}_{x+\nu,\mu} U^{-1}_{x,\nu} + U_{x,\mu} \delta U_{x+\mu, \nu} U^{-1}_{x+\nu,\mu} U^{-1}_{x,\nu}\notag \\
            &+ U_{x,\mu} U_{x+\mu, \nu} \delta U^{-1}_{x+\nu,\mu} U^{-1}_{x,\nu} + U_{x,\mu} U_{x+\mu, \nu} U^{-1}_{x+\nu,\mu} \delta U^{-1}_{x,\nu}.
        \end{align}
        We then reorder terms in the trace and rename indices to write the variation as
        \begin{align}
            \delta S_\mathrm{W}[U, \delta U] = \frac{1}{2N_c} \sum_{x,\mu\neq\nu} i \delta \tilde A^a_{x,\mu} \mathrm{Tr} \left[ t^a \left( \beta_{x,\mu\nu} \left( U_{x,\mu\nu} - U^{-1}_{x,\mu\nu} \right) + \beta_{x-\nu,\mu\nu} \left( U_{x,\mu-\nu} - U^{-1}_{x,\mu-\nu} \right) \right)\right],
        \end{align}
        which allows us to read off the drift term
        \begin{align}
            \frac{\delta S_\mathrm{W}}{\delta \tilde A^a_{x,\mu}} = \frac{i}{2N_c} \sum_\nu \mathrm{Tr} \left[ t^a \left( \beta_{x,\mu\nu} \left(U_{x,\mu\nu} - U^{-1}_{x,\mu\nu} \right) + \beta_{x-\nu, \mu\nu} \left(U_{x,\mu-\nu} - U^{-1}_{x,\mu-\nu} \right) \right) \right].
        \end{align}
        Inserting the definition of the coupling constants $\beta_{x,\mu\nu}$, we find that the components of the Wilson drift term are given by
        \begin{align}
        \label{eq:dSW_dAtilde}
        \frac{\delta S_\mathrm{W}}{\delta \tilde A^a_{x,t}} &= - \frac{i}{2N_c} \beta_{k, 0} \sum_{|i|} \mathrm{Tr} \left[ t^a \left(U_{x,0i} - U^{-1}_{x,0i} \right) \right], \\
        \frac{\delta S_\mathrm{W}}{\delta \tilde A^a_{x,i}} &= - \frac{i}{2N_c}  \mathrm{Tr} \left[ t^a \left( \beta_{k,0} \left(U_{x,i0} - U^{-1}_{x,i0} \right) + \beta_{k-1,0} \left(U_{x,i-0} - U^{-1}_{x,i-0}\right) \right)\right] \notag \\
        &\quad + \frac{i}{2N_c} \beta_{k, s} \sum_{|j|} \mathrm{Tr} \left[ t^a \left(U_{x,ij} - U^{-1}_{x,ij} \right) \right].
        \end{align}
        The sums over $|i|$ and $|j|$ run over positive and negative orientations of the direction, i.e. given some arbitrary expression $X_i$ 
        \begin{align}
            \sum_{|i|} X_i \equiv \sum^3_{i=1} \left( X_i + X_{-i} \right).
        \end{align}
        Note that the drift terms computed from the variation coincide with the group derivative used in \cite{Berges:2006xc}
        \begin{align} \label{eq:group_derivative}
            D^a_{x,\mu} S_\mathrm{W}[U] \equiv \lim_{\varepsilon \rightarrow 0 } \frac{S_\mathrm{W}[e^{i \varepsilon t^a} U_{x,\mu}] - S_\mathrm{W}[U]}{\varepsilon} = \frac{\delta S_\mathrm{W}}{\delta \tilde A^a_{x,\mu}}.
        \end{align}
        In the above expression, a single link $U_{x,\mu}$ of the set of gauge links is perturbed by a group element $e^{i \varepsilon t^a}$ close to the unit element. 
        
        In order to identify the continuum drift term with an appropriate approximation in terms of the Wilson action, we discretize the variation of the Yang-Mills action on the lattice via
        \begin{align} \label{eq:ym_variation_approx}
            \delta S_\mathrm{YM}[A_\mu, \delta A_\mu] &\approx \sum_{k, \mx}  \left( a_{t,k} a^3_s\, \frac{\delta S_\mathrm{YM}}{\delta A^a_t(t_k+a_{t,k}/2, \mx)}\, \delta A^a_t \left(t_k+\frac{a_{t,k}}{2}, \mx \right) \right.\nonumber 
            \\ &\qquad\, + \left.\bar a_{t,k} a^3_s\, \frac{\delta S_\mathrm{YM}}{\delta A^a_i(t_k, \mx + a_s \hat e_i /2)}\, \delta A^a_i \left(t_k, \mx + \frac{a_s}{2} \hat e_i \right) \right),
        \end{align}
        where we have approximated the volume element in the spatial part using the averaged time-step $\bar a_{t,k}$. Comparing the above expression to \eq\eqref{eq:wilson_variation} and accounting for the absorbed factors $g a_\mu$ in $\delta \tilde A^a_{x,\mu}$ allows us to relate the Yang-Mills drift to the Wilson drift term via
        \begin{align}\label{eq:drift_identification}
            \frac{\delta S_\mathrm{YM}}{\delta A^a_t(t_k+a_{t,k}/2, \mx)} \approx \frac{g}{a_s^3} \frac{\delta S_\mathrm{W}}{\delta \tilde A^a_{x,t}}, \qquad
            \frac{\delta S_\mathrm{YM}}{\delta A^a_i(t_k, \mx + a_s \hat e_i /2)} \approx \frac{g}{\bar a_{t,k} a_s^2}  \frac{\delta S_\mathrm{W}}{\delta \tilde A^a_{x,i}}.
        \end{align}

        \subsection{Deriving the CL equation for gauge links} 
        
        Having worked out the details regarding the discrete action and how its drift term is related to the theory in the continuum, we can now perform the discretization of the contour-parameterized CL equation for gauge fields. For practical purposes, we start from \eq\eqref{eq:cle_ym_contour_hybrid} in which the gauge fields are given in the $(t, \mx)$ frame, whereas the noise is still in the $(\lambda, \mx)$ formulation. For better readability, we restate this evolution equation here:
        \begin{align}
            \begin{split}
                \p_\theta A^a_t(\theta, t, \mx) &= i \,\frac{d\lambda}{dt}\left.\frac{\dd S_\mathrm{YM}}{\dd A^a_t(t, \mx)}\right\vert_\theta + \frac{d\lambda}{dt}\,\eta^a_\lambda(\theta, \lambda, \mx), \\
                \p_\theta A^a_i(\theta, t, \mx) &= i\,\frac{dt}{d\lambda} \left.\frac{\dd S_\mathrm{YM}}{\dd A^a_i(t, \mx)}\right\vert_\theta + \eta^a_i(\theta, \lambda, \mx).
            \end{split}
        \end{align}
        In analogy to \app\ref{app:disc_drift}, we discretize the Langevin time $\theta$ in steps of $\epsilon$ and absorb a factor of $\sqrt{\epsilon}$ into the definition of the noise term. 
        For consistency with \eq\eqref{eq:link_continuum}, we approximate the gauge fields $A^a_\mu(t, \mx)$ and noise fields $\eta^a_\mu(\lambda, \mx)$ at the mid-points of edges $x+\hat\mu/2$. The derivative terms $d\lambda/dt$ and $dt/d\lambda$ appearing in front of the drift terms and the noise fields are approximated with appropriate central finite differences. For the temporal update step we use
        \begin{align}
            \frac{d\lambda}{dt} \approx \frac{\lambda_{k+1} - \lambda_k}{t_{k+1} - t_k} = \frac{a_{\lambda, k}}{a_{t,k}}
        \end{align}
        and 
        \begin{align}
            \frac{dt}{d\lambda} \approx \frac{t_{k+1} - t_{k-1}}{\lambda_{k+1} - \lambda_{k-1}} = \frac{\bar a_{t, k}}{\bar a_{\lambda,k}}
        \end{align}
        for the spatial update step, which conserves time reversal symmetry with the averaged time-step \eqref{eq:avg_timestep}. Employing \eq\eqref{eq:drift_identification} then yields
        \begin{align}
            \begin{split}
            A^a_{x,t}(\theta+\epsilon) &= A^a_{x,t}(\theta) + i \epsilon \frac{a_{\lambda,k}}{a_{t,k}}\, \frac{g}{ a_s^3} \left.\frac{\delta S_\mathrm{W}}{\delta \tilde A^a_{x,t}} \right\vert_\theta + \sqrt{\epsilon}\, \frac{a_{\lambda,k}}{a_{t,k}} \eta^a_{x,\lambda}(\theta), \\
            A^a_{x,i}(\theta+\epsilon) &= A^a_{x,i}(\theta) + i \epsilon \frac{\bar a_{t,k}}{ \bar a_{\lambda,k}}\, \frac{g}{\bar a_{t,k} a_s^2}  \left. \frac{\delta S_\mathrm{W}}{\delta \tilde A^a_{x,i}} \right\vert_\theta + \sqrt{\epsilon}\, \eta^a_{x,i}(\theta).
        \end{split}
    \end{align}
    The symmetrically discretized noise correlators that have absorbed factors of $\sqrt{\epsilon}$ read
    \begin{align}
        \langle \eta^a_{x,\lambda}(\theta) \eta^b_{x',\lambda}(\theta') \rangle &= 2 \delta^{ab} \delta_{\theta\theta'} \frac{1}{a_{\lambda,k} a_s^3} \delta_{kk'} \delta_{\mx, \mx'}, \\
        \langle \eta^a_{x,i}(\theta) \eta^b_{x',j}(\theta') \rangle &= 2 \delta_{ij} \delta^{ab} \delta_{\theta\theta'} \frac{1}{\bar a_{\lambda,k} a_s^3} \delta_{kk'} \delta_{\mx, \mx'}.
    \end{align}
    Depending on the component of the noise field, we have chosen symmetric discretizations of the Dirac distribution $\delta(\lambda - \lambda')$, resulting in factors of either $a_{\lambda,k}$ or averaged steps $\bar a_{\lambda,k}$ in the correlators.
    
    The next step is to rewrite the discrete CL equations in terms of dimensionless degrees of freedom. This is done by absorbing factors of $g a_{t,k}$ and $g a_s$ into the definition of the gauge fields $\tilde A^a_{x,\mu} = g a_\mu A^a_{x,\mu}$
    and also redefining the noise fields as $\eta^a_{x,\lambda} \rightarrow \eta^a_{x,\lambda} / \sqrt{a_{\lambda,k} a_s^3}$ and $\eta^a_{x,i} \rightarrow \eta^a_{x,i} / \sqrt{\bar a_{\lambda,k} a_s^3}$, which leads to a dimensionless noise correlator
    \begin{align} \label{eq:ym_cle_correlator_app}
        \langle \eta^a_{x,\mu}(\theta) \eta^b_{x',\nu}(\theta') \rangle &= 2 \delta_{\mu\nu} \delta^{ab} \delta_{\theta\theta'} \delta_{kk'} \delta_{\mx, \mx'}.
    \end{align}
    The resulting CL equations read
    \begin{align}
        \tilde A^a_{x,t}(\tilde \theta+\tilde \epsilon) &= \tilde A^a_{x,t}(\tilde \theta) + i \tilde \epsilon\, \frac{a_{\lambda,k}}{a_s}  \left.\frac{\delta S_\mathrm{W}}{\delta \tilde A^a_{x,t}} \right\vert_{\tilde \theta} + \sqrt{\tilde \epsilon} \sqrt{\frac{a_{\lambda,k}}{a_s}} \, \eta^a_{x,\lambda}(\tilde \theta), \\
        \tilde A^a_{x,i}(\tilde \theta+ \tilde \epsilon) &= \tilde A^a_{x,i}(\tilde \theta) + i \tilde \epsilon\, \frac{a_s}{\bar a_{\lambda,k}}   \left. \frac{\delta S_\mathrm{W}}{\delta \tilde A^a_{x,i}} \right\vert_{\tilde \theta} + \sqrt{\tilde \epsilon} \sqrt{\frac{a_s}{\bar a_{\lambda,k}}}\, \eta^a_{x,i}(\tilde \theta),
    \end{align}
    where we have also performed a redefinition of the Langevin time $\theta \rightarrow \tilde \theta = g^2 \theta / a_s^2$ with Langevin step $\tilde \epsilon = g^2 \epsilon / a_s^2$. For notational simplicity, we drop the tilde in the Langevin step for the rest of this work.
    
    Finally, we formulate the above equations in terms of gauge links by multiplying by $it^a$ and exponentiating both sides. Making use of the continuum limit approximation in \eq\eqref{eq:link_continuum} then yields CL equations in terms of gauge links. In this last step, we demand the Langevin step $\epsilon$, \new{and the lattice spacings $a_s$ and $a_{\lambda, k}$} to be sufficiently small such that we can neglect higher order terms from the Baker-Campbell-Hausdorff formula. 
    Our final result reads
    \begin{align} \label{eq:ym_contour_cle_unkerneled_temp_app}
        U_{x,t}( \theta +  \epsilon) &= \exp \left( i t^a \left[ i  \epsilon\, \frac{a_{\lambda,k}}{a_s}  \left.\frac{\delta S_\mathrm{W}}{\delta  \tilde A^a_{x,t}} \right\vert_{ \theta} + \sqrt{ \epsilon}\, \sqrt{\frac{a_{\lambda,k}}{a_s}}\, \eta^a_{x,\lambda}( \theta) \right] \right) U_{x,t}( \theta), \\ \label{eq:ym_contour_cle_unkerneled_spat_app}
        U_{x,i}( \theta +  \epsilon) &=  \exp \left( i t^a \left[ i  \epsilon\, \frac{a_s}{\bar a_{\lambda,k}}   \left. \frac{\delta S_\mathrm{W}}{\delta \tilde A^a_{x,i}} \right\vert_{ \theta} + \sqrt{ \epsilon} \, \sqrt{\frac{a_s}{\bar a_{\lambda,k}}}\, \eta^a_{x,i}( \theta) \right] \right) U_{x,i}( \theta).
    \end{align}

\section{Stabilization techniques} \label{sec:stabilization}
    It is known that CL simulations for complex actions primarily suffer from two types of instabilities. One of these is that the stochastic process may diverge, which leads to a breakdown of the simulation as the drift term blows up. We refer to this issue as runaway instability. Furthermore, the Langevin process can approach wrong stationary solutions, which is known as \emph{wrong convergence} (see e.g.~\cite{Berges:2006xc} where this occurs for real-time lattice gauge theories). There may be two root causes for wrong convergence. Firstly, the spectrum of the corresponding Fokker-Planck (FP) operator is required to be negative semi-definite. Otherwise there is no guarantee that the stationary solution describes the desired path integral.
    Secondly, a wrong stationary solution may result from non-vanishing boundary terms that spoil the criterion of correctness in the derivation of the CL method \cite{Nagata:2016vkn, Scherzer:2018hid, Scherzer:2018udt}. Hence, the issues of wrong convergence of the stochastic process may in some cases not be primarily of numerical nature, i.e., due to the discretization of the Langevin equation,%
    \footnote{Nevertheless, the choice of the numerical scheme for the discrete Langevin equation can be highly important for stability \cite{Alvestad:2021hsi}.} 
    but rather of a more fundamental origin \cite{Aarts:2009uq}.
    For lattice gauge theories, some of these issues can be mitigated using modern stabilization techniques, which we summarize in this part of the appendix.
    
    \paragraph{Adaptive step size (AS)}
    
        The run away instability can be removed by the introduction of an adaptive Langevin time step $\epsilon$ \cite{Aarts:2009dg}. We modify $\epsilon$ by introducing an upper bound $B$ of the drift term $K_{x,\mu}^a = \delta S_\mathrm{W} / \delta \tilde A^a_{x,\mu}$ by reducing the step size if the latter becomes too large. This is done with the substitution
        \begin{align} \label{eq:as}
            \epsilon \mapsto \tilde \epsilon = 
            \epsilon\; \min\left(1, \frac{B}{\max\limits_{x,\mu, a} \vert K_{x,\mu}^a\vert} \right).
        \end{align}
        We set the bound parameter $B$ small enough that it regularizes potentially large drift terms, which are responsible for runaway instabilities. Moreover, $B$ is tuned in such a way that it is sufficiently larger than the average of the maximal 
        drift term in the metastable region in order to avoid large biases of the simulation results.
    
    \paragraph{Gauge cooling (GC)}
        
        It has been empirically shown that the minimization of a functional known as the unitarity norm $F[U]$  can mitigate wrong convergence instabilities. The unitarity norm measures the non-unitarity of the link configuration, or more intuitively, it measures how ``far away'' the complexified links are from the unitary subgroup. Gauge cooling exploits the gauge freedom of the complexified system under {SL($N_c$, $\mathbb C$)} transformations, which allows us to minimize the unitarity norm using gauge transformations. 
        For the minimization, we adopt the gauge cooling procedure introduced in \cite{Seiler:2012wz} and developed further in \cite{Aarts:2013uxa}.

        As a non-unitarity measure we use a version of the unitarity norm
        \begin{align} \label{eq:unorm}
            F[U] = \sum\limits_{x,\mu} \mathrm{Tr}\left[(U_{x, \mu} U_{x, \mu}^\dagger - 1)^2\right],
        \end{align}
        which differs from the original formulation by the inclusion of the square. We found that this modification turns out to be more efficient in our simulations. The minimization process is done by gauge transforming the links according to
        \begin{align} \label{eq:gc}
            U_{x,\mu} \; \mapsto \; &U_{x}^V = V_{x,\mu} U_{x, \mu} V_{x+\mu}^{-1}, \qquad F[U] \geq F[U^V],
        \end{align}
        where the gauge transformation $V_{x} = \exp (i\,\alpha_{\mathrm{GC}}\, \gamma_x)$ (with $\gamma_x = t^a \gamma_x^a \in \mathfrak{su}(N, \mathbb C)$) is determined by a gradient descent scheme. The gauge gradient is calculated via (see also \eq\eqref{eq:group_derivative})
        \begin{align}
            D_{x,\mu}^{a} f[U] = \lim_{\epsilon\to 0} \frac{1}{\epsilon} \Big[ f[U'(\delta)] - f[U] \Big],
        \end{align}
        where $f[U'(\delta)]$ is given by the functional evaluated for the modified link configuration $U'$,
        which is altered at $x$ and $\mu$ such that $U_{x,\mu}$ is replaced by $U'_{x,\mu} = \exp\left(i \epsilon t^a\right)\, U_{x,\mu}$. 
        
        We obtain a suitable gauge transformation by computing the gauge gradient of $F[U^V]$ and choosing $V$ such that the gradient points towards the steepest decent. We find
        \begin{align}
            \gamma_x^a = 
            &- \sum_\mu \Tr
            \left[
                t^a \left( W_{x}^+ - W_{x}^-\right)
            \right],
        \end{align}
        where we use
        \begin{align}
            W_{x}^+ &= \sum_\mu \left( U_{x,\mu} U_{x,\mu}^\dagger - 1 \right) U_{x,\mu} U_{x,\mu}^\dagger , \\
            W_{x}^- &= \sum_\mu \left( U_{x-\mu,\mu}^\dagger U_{x-\mu,\mu} - 1 \right) U_{x-\mu,\mu}^\dagger U_{x-\mu,\mu}.
        \end{align}
        Gauge transforming the link configuration during the CL simulation amounts to adding another term to the FP operator of the FP equation. In \cite{Nagata:2015uga} it was shown that this additional term vanishes when applied on gauge invariant observables and does not bias the results.
        
        Gauge cooling is iteratively performed after each CL update step. The number of gauge cooling steps and the magnitude of $\alpha_{\mathrm{GC}}$ depend on the lattice size, the initial configuration and the Langevin time step of the simulation. In \cite{Aarts:2013uxa} it was further observed that gauge cooling converges slower for larger lattices, and adaptive GC steps were introduced. However, we find that these adaptive techniques do not have a significant advantage for the purposes of our real-time Yang-Mills simulations.

    \paragraph{Dynamical stabilization (DS)}

        As a third stabilization technique we use dynamical stabilization \cite{Attanasio:2018rtq}. It introduces an additional term in the CL update equation that counteracts large imaginary drifts. 
        The drift term is substituted by
        \begin{align}
            &K_{x,\mu}^a \mapsto \tilde{K}_{x,\mu}^a = K_{x,\mu}^a + i \alpha_\mathrm{DS} M_{x}^a, \\
            &M_{x}^a = b_{x}^a \left( \sum_{c} b_{x}^c b_{x}^c \right), \quad 
            b_{x}^a = \sum_{\mu} \mathrm{Tr}[t^a U_{x,\mu} U_{x,\mu}^\dagger],
        \end{align}
        where the penalty term $M_x^a$ acts on the imaginary part of the drift term and is proportional to the local non-unitary of the configuration. This leads to a reduction of the unitarity norm, but can not be understood as an admissible gauge transformation. Nevertheless, this method has resulted in practical advancements in the calculation of the QCD equation of state at finite density \cite{Attanasio:2022mjd}. In contrast to gauge cooling, DS has (as of yet) no rigorous justification.
        
        The penalty term is a non-physical modification of the CL update step. Thus, we have to choose the force parameter $\alpha_\mathrm{DS}$ cautiously.  In practice,  $\alpha_\mathrm{DS}$ needs to be tuned to minimize possible biases of measured observables, while at the same time improving the stability of the simulation. In our applications we observed that the bias becomes significant when the penalty term is of the same order of magnitude as the drift term.
        We found that DS impacts our results negatively when the unitarity norm grows too quickly.

\bibliographystyle{JHEP}
\bibliography{clpaper}

\end{document}